\documentclass{aa}  
\makeatletter
\renewcommand*\aa@pageof{, page \thepage{} of \pageref*{LastPage}}
\makeatother
\usepackage{graphicx}
\usepackage{txfonts}
\usepackage[dvipsnames]{xcolor}
\usepackage[colorlinks=true, allcolors=blue]{hyperref}
\begin{document}

   \title{Star formation histories of Coma cluster galaxies matched to simulated orbits hint at quenching around first pericenter}

   \author{A.~K.~Upadhyay 
          \inst{1},
          K.~A.~Oman
          \inst{1,2},
          \and
          S.~C.~Trager\inst{1}
          }
   \institute{Kapteyn Astronomical Institute, University of Groningen,
              Landleven 12, 9747 AD Groningen\\
              \email{upadhyay@astro.rug.nl}\\
              \email{sctrager@astro.rug.nl}
         \and
             Durham University, Department of Physics - Astronomy, South Road, DH1 3L Durham\\
             \email{kyle.a.oman@durham.ac.uk}
             }

    \titlerunning{SFH of Coma Cluster galaxies matched to simulated orbits}
    \authorrunning{Upadhyay, Oman and Trager}

  \abstract
   {The star formation in galaxies in present-day clusters has almost entirely been shut down, but the exact mechanism that quenched these galaxies is still uncertain.}
   {By tracing the orbital and star formation histories of galaxies within the Coma cluster, we seek to understand the role of the high-density cluster environment in quenching these galaxies.}
   {We combine star formation histories extracted from high-signal-to-noise spectra of 11 early-type galaxies around the center of the Coma cluster with probability distributions for their orbital parameters obtained using an N-body simulation to connect their orbital and star formation histories.}
   {We find that all 11 galaxies likely quenched near their first pericentric approach. Higher stellar mass galaxies ($\log(M_\star/\mathrm{M}_\odot)>10$) had formed a higher fraction of their stellar mass (more than $\sim90\%$) than their lower mass counterparts ($\sim 80$--$90\%$) by the time they fell into the cluster (when they cross $2.5r_\mathrm{vir}$). We find that the expected infall occurred around $z\sim0.6$, followed by the first pericentric passage $\sim4\,\mathrm{Gyr}$ later. Galaxies in our sample formed a significant fraction of their stellar mass, up to $15\%$, between infall and first pericenter, and had assembled more than $\sim98\%$ of their cumulative stellar mass by first pericenter.}
   {Unlike previous low-redshift studies that suggest that star formation continues until about first apocenter or later, the high percentage of stellar mass already formed by first pericenter in our sample galaxies points to star formation ceasing within a gigayear after the first pericentric passage. We consider the possible physical mechanisms driving quenching and find that our results resemble the situation in clusters at $z\sim1$, where active stripping of gas (ram-pressure or tidally driven) seems to be required to quench satellites by their first pericentric passage. However, a larger sample will be required to conclusively account for the unknown fraction of preprocessed satellites in the Coma cluster.}


   \maketitle
%

\section{Introduction}\label{sec:intro}

   Galaxies occupy different positions in the cosmic web, and their observable properties -- such as luminosity, color, morphology, and gas content -- at the current epoch are direct results of their evolutionary histories in different environments. The distribution of galaxies in the local Universe shows a bimodality between ``blue-cloud'' and ``red-sequence'' galaxies \citep[e.g.,][]{Strateva2001,Baldry_2004,Baldry_2006}. Blue-cloud galaxies are typically gas rich and star forming, with late-type morphologies, while red-sequence galaxies are typically gas poor with little or no star formation and exhibit early-type morphologies. The fraction of red galaxies increases with both galaxy mass and environmental density \citep[e.g.,][]{Dressler1980,Kauffmann2004,Baldry_2004,Baldry_2006}. Galaxies in the high-density environment of clusters are therefore more likely to be quiescent as compared with their similar-mass counterparts in the field \citep{Hogg2004}, an observation further supported by several studies that have found cluster galaxies to be relatively atomic gas deficient \citep[e.g.,][]{gavazzi198721,fumagalli2009molecular,boselli2014cold}. The density dependence is significant both within a cluster and out to several times the nominal radius of such systems because the fraction of quiescent galaxies increases with the increase in density toward the cluster core \citep[e.g.,][]{Balogh2000,Lewis2002,Gomez2003}.
   
   Within a cluster, there are various mechanisms at play that lead to the quenching\footnote{We adopt the definition of, e.g., \citet[][]{de2012environmental} and \citet[][]{Wetzel_2013} in defining a threshold specific star-formation rate (SSFR) separating active and quiescent (or ``quenched'') galaxies at $10^{-11}\,\mathrm{yr}^{-1}$.} of galaxies. In addition to ``mass quenching,'' which affects all galaxies irrespective of their local environment \citep[e.g.,][]{Larson1974,Binney1977,Dekel1986,Birnboim2003,Croton2006,Martig_2009,fabian2012,cicone2014}, cluster galaxies are also subject to ``environmental quenching.'' \citet{Peng_2010} have shown that environmental and mass quenching efficiency are (nearly) independent. Lower mass galaxies are typically star forming, unless they are quenched by their environment, while massive galaxies are more prone to being quenched regardless of their environment. We now discuss various quenching mechanisms.

   There are two major types of environmental quenching, depending on the type of interaction: One is due to the interaction between gas in the interstellar medium (ISM) of a galaxy and the hot gas of the intra-cluster (or intra-group) medium (ICM), and the other is due to the gravitational interaction of galaxies with other cluster members, including the (typically central) brightest cluster galaxy. Effective galaxy-galaxy interactions are rare in clusters due to the high relative speeds of satellites \citep[see][Ch.~8, especially Sect.~8.2g]{binney_tremaine_2008}, but repeated high-speed encounters between satellites, termed ``galaxy harassment,'' can dynamically heat them, either directly removing some gas or rendering it more susceptible to other stripping processes via gravitational heating \citep{Moore1996,Moore1998}. Tidal effects due to the cluster itself are strongest near its core \citep{merritt1983,Mayer2006}, where satellites can be tidally stripped of gas, stars, and dark matter.
   
   Some satellite galaxies enter their host cluster more susceptible to quenching because the environment of a previous, lower mass host has caused a decline in their star formation. These satellites are said to be preprocessed \citep{balogh1999,poggianti1999,de2012environmental,Wetzel_2013,bahe2013,taranu2014,Roberts_2017,haines2018,Smith_2019,pallero2019,pasquali2019,Rhee_2020}. Several studies have found a high fraction ($\sim 50$--$65\%$) of quiescent galaxies in the immediate surroundings of clusters, which may be attributable to preprocessing \citep{Wetzel_2013,bahe2013,taranu2014,haines2018,pallero2019}. The hierarchical nature of the fiducial $\Lambda$CDM cosmological model means that the environments around massive clusters preferentially contain smaller groups of galaxies \citep[see Fig.~1 of][]{Wetzel_2013}, so preprocessing plays a significant role in the observed quiescent fraction in and around clusters. The fraction of preprocessed galaxies rises with the final cluster mass, and therefore group preprocessing is more prominent in the satellites of massive ($\log(M_\star/\mathrm{M}_\odot) > 14.5$) clusters, such as the Coma cluster \citep{Hou_2014,Roberts_2017,Smith_2019,pallero2019}.
   
   Apart from the group preprocessed population in the outskirts of the cluster, the rest of the satellite population starts experiencing environmental effects even before entering the cluster virial\footnote{We define virial quantities following \cite{Bryan1998}: The virial radius at $z=0$ encloses a spherical volume within which the mean density is $\approx 360$ times the mean matter density of the Universe. Many studies instead use a virial overdensity of $200$ times the critical density for closure; an approximate conversion, valid at $z\sim0$, is given by $r_\mathrm{200c}/r_\mathrm{360b} \sim 0.73$, where the $b$ in $r_\mathrm{360b}$ stands for ``background,'' i.e., $360\Omega_\mathrm{m}\rho_\mathrm{crit}$.} radius. Accretion of fresh gas becomes inefficient, and the satellites' hot coronal gas -- which acts as a fuel source for star formation by replenishing the cold gas disk -- is slowly stripped away, causing a ``strangulation'' \citep{McGee_2014} of the gas supply required for star formation.
   
   The deep gravitational potential well of a cluster results in high orbital velocities of its satellites and heats up the gas in the ICM. As the cluster galaxies pass through the hot ICM at high speed, their neutral and atomic gas collides with the ICM and may be removed via ram-pressure stripping \citep[RPS;][]{Gunn1972,abadi1999ram,jachym2007}. Ram-pressure stripping likely plays a key role in the quenching of satellite galaxies \citep{Jaffe_2015}. Since it is proportional to the density of the ICM, and to the square of the orbital speed, ram pressure peaks strongly around a satellite's pericenter, though in some cases it can effectively strip their ISM out to at least the virial radius \citep{tonnesen2007}. \citet{roberts2019quenching} found that an increase in the quenched fraction near the centers of clusters ($r<0.25r_\mathrm{vir}$) can be attributed to RPS. Several other studies \citep[e.g.,][]{Jaffe_2018,lotz2019gone,maier2019cluster,roberts2020ram,Rhee_2020,Oman2021} also suggest that RPS plays an important role in satellite quenching.


    Several quenching processes act on cluster satellites, and their strengths are dependent on the cluster-centric distance of the satellite and the time spent inside the cluster. One possible approach to constrain the dominant processes is to establish the dependence of their star formation histories (SFHs) on their orbital histories \citep[e.g.,][]{Wetzel_2013,Oman_2016,Rhee_2020,Oman2021}. 
   
   The orbital trajectories and phases of satellites are not directly observable but can be constrained using their projected cluster-centric radius and velocity offset along the line of sight: concisely, their ``projected phase space'' (PPS) coordinates. Using simulations, galaxy orbits around clusters can be traced backward/forward in time from their current position in PPS \citep[e.g.,][]{Mamon_2004,Mahajan2011,Oman_2013,Wetzel_2013,Oman_2016,pasquali2019,Rhee_2020,Oman2021}. The distribution of galaxies in and around clusters can be loosely decomposed into three populations: infall, backsplash, and virialized. Infalling satellites are on their initial approach toward their host cluster; virialized galaxies have settled into its central regions; and the backsplash population consists of satellites on eccentric orbits and near their first apocenter, after an initial orbit through the host. The mapping between PPS and orbital phase is not one-to-one: For instance, the PPS location of the backsplash population partially overlaps that of the infalling population. This overlap can make it difficult to distinguish between backsplash galaxies quenched by the cluster, and preprocessed infalling galaxies.
   
   Previous studies have used different parameters to describe the orbital histories of satellites. Examples include the time since infall \citep[TSI; e.g.,][]{Wetzel_2013,Oman_2016,Rhee_2020}, the time since (or until) first pericenter \citep{Oman2021}, and dividing the PPS into zones respectively containing primarily recent and ancient infallers \citep{Mahajan2011,Hou_2014,noble2016,pasquali2019,Smith_2019}.


    To obtain a better understanding of the physics of quenching it is also essential to examine how the quenching timescale ($t_\mathrm{q}$), gas depletion timescale ($t_\mathrm{depl}$), and dynamical timescale ($t_\mathrm{dyn}$) evolve with redshift. The dynamical timescale -- or crossing time -- is likely a key driver of the redshift evolution of the quenching timescale \citep[e.g.,][]{McGee_2014,foltz2018evolution,Rhee_2020}. It is proportional to $(1+z)^{-1.5}$, or nearly linearly proportional to lookback time. The gas depletion timescale describes the time it would take to use up all the gas to form stars at current star-formation rate (SFR) if the gas is not removed earlier by some other process. It plays a critical role in regulating the gas reservoir for star formation. Using the xCOLD GASS \citep{xcoldgass} and xGASS \citep{xgass} surveys that provide a census of molecular and neutral gas in the galaxies of the local Universe, \citet{Feldmann2020} found that the scaling relation of SFR and total gas mass (molecular+neutral) is directly linked to the gas depletion timescale \citep[see also][for the redshift dependence of the gas depletion timescale]{tacconi2018phibss}. Cosmological simulations show that satellite halos stop accreting substantial amounts of mass when they are inside the virialized region of their host halo \citep[e.g.,][]{mergertree}, so when satellites are accreted onto a cluster, the supply of fresh gas is cut off around the time of infall. After this, satellites should continue to form stars from their limited cold gas supply, and hot gas reservoir provided it can cool, unless these are removed from one or many of the gas removal processes discussed in Sect.~\ref{sec:intro}.
    
    Several studies have found longer $t_\mathrm{q}$ in clusters at lower redshifts. For example, \citet{Wetzel_2013}, \citet{taranu2014}, and \citet{balogh2016evidence} found $t_\mathrm{q} \sim 4.4 \pm 0.4$\,Gyr, $4 \pm 2$\, Gyr, and $5 \pm 0.5$\,Gyr, respectively, at $z \sim 0$. \citet{Haines_2015} found $t_\mathrm{q} \sim 3.7$\,Gyr at $z \sim 0.2$. At $z \sim 1$, \citet{Muzzin_2014,balogh2016evidence,foltz2018evolution} found $t_\mathrm{q} \sim 1 \pm 0.25$\,Gyr, $1.5 \pm 0.5$\,Gyr, and $1.3 \pm 0.5$\,Gyr, respectively. \citet{foltz2018evolution} studied quenching timescales at an even higher redshift of $z \sim 1.5$, finding $t_\mathrm{q} \sim 1.1 \pm 0.5$\,Gyr. The longer $t_\mathrm{q}$ found by low-redshift studies is suggestive of quenching well after the first pericentric passage \citep[see also][]{Oman2021}, while the shorter timescales at higher redshifts are more consistent with quenching around the first pericenter.

   In this study we explore the effects of environment on cluster galaxies. We attempt to constrain the physics driving quenching by comparing the SFHs of satellites with their probable orbital histories. We extract stellar properties from high-S/N spectra of 11 Coma cluster galaxies, and constrain their orbits by comparing with a library compiled from the VVV-L0 N-body simulation extended into the future, up to $z = -0.5$ ($\approx 10$\,Gyr from now). In contrast with other studies, we parametrize the orbit by both the expected infall time (when galaxies fall within $2.5r_\mathrm{vir}$ of the cluster) and the time of first pericenter. By determining the fraction of their final stellar mass assembled by the expected infall and first pericenter times, we reconstruct a simple description of the linked orbital and SFHs of our sample.
 
\section{Data and methods}
    We present the SFHs of our sample of early-type Coma satellites in Sect.~\ref{sec:stellarpops}, and the libraries of satellite orbits drawn from an N-body simulation in Sect.~\ref{sec:orbits}, and how we use these to estimate the orbital parameters of observed satellites in Sect.~\ref{sec:obsorbits}.
    
    \begin{table*}
    \centering
        \caption[]{Properties of Coma cluster galaxies.}
         \label{tab:coma_prop1}
        \begin{tabular}{l l r r c l r}
        \hline
        \noalign{\smallskip}
        GMP~& Other name(s) & RA (J2000) & Dec. (J2000) & $z$ & Morphology & \multicolumn{1}{c}{$\log(M_\star / \mathrm{M}_\odot)$} \\
        \noalign{\smallskip}
        \hline
        \noalign{\smallskip} 
        3254 & D127, RB042 & 12:59:40.3 & +27:58:06 & 0.02512 & S0 & 9.92\\ 
        3269 & D128, RB040 & 12:59:39.7 & +27:57:14 & 0.02678 & S0 & 9.98\\ 
        3291 & D154, RB038 & 12:59:38.3 & +27:59:15 & 0.02260 & S0 & 9.92\\ 
        3352 &  NGC~4872 & 12:59:34.2 & +27:56:48 & 0.02007 & SB0 & 10.62\\
        3367 &  NGC~4873 & 12:59:32.7 & +27:59:01 & 0.02403 & S0 & 10.65\\ 
        3414 &  NGC~4871 & 12:59:30.0 & +27:57:22 & 0.02253 & SB0 & 10.73\\
        3484 & D157, RB014 & 12:59:25.5 & +27:58:23 & 0.01596 & Sa & 10.11\\ 
        3534 & D158, RB007 & 12:59:21.5 & +27:58:25 & 0.02244 & S & 9.67\\ 
        3565 & RB005 & 12:59:19.8 & +27:58:26 & 0.02399 & E & 9.29\\ 
        3639 & NGC~4867 & 12:59:15.2 & +27:58:16 & 0.01931 & E5 & 10.60\\
        3664 & NGC~4864 & 12:59:13.1 & +27:58:38 & 0.02393 & E1 & 10.82\\ \noalign{\smallskip}
        \hline
        \end{tabular}
    \tablefoot{The general properties of observed galaxies are taken from \citet{Trager_2008}, except for the redshifts, which were obtained from the observed spectra using STECKMAP. Morphologies were taken from the following sources: GMP~3254, 3269, 3484, 3534, 3639, 3664 from \citet{michard2008}; GMP~3291, 3352, 3367, 3414 from \citet{Lansbury2014}; and GMP~3565 from \citet{eisenhardt2007}. The other names of our sample galaxies are taken from ``RB''/[RB67] catalog \citep{rood1967} and ``D''/[D80] catalog \citep{Dressler1980}.}
    \end{table*}

\subsection{STECKMAP star formation histories}\label{sec:stellarpops}    
    We use spectra from \citet{Trager_2008}, who analyzed the stellar population histories of 12 early-type galaxies (ETGs) in the Coma cluster. The spectra were taken on 7 April 1997 and consist of three consecutive 30-minute exposures taken with the Low Resolution Imaging Spectrograph \citep[LRIS;][]{oke1995} on the Keck-II telescope. The spectra have wavelength coverage between $3500$--$6000$\,\AA\ with a spectral resolution of $4.4$\,\AA\ (FWHM) and signal-to-noise (S/N) values ranging between $100$--$300$, except for GMP~3291, GMP~3534, and GMP~3565, which have slightly lower S/N values of 59, 82, and 35, respectively. General properties of our galaxy sample are listed in Table~\ref{tab:coma_prop1}. We ignore GMP~3329 in this work as it is a central and not a satellite galaxy in the Coma cluster.

    We use STECKMAP \citep[STEllar Content and Kinematics via Maximum A Posteriori,][]{Ocvirk_2006} to extract the stellar and kinematic properties of our galaxy sample.
    STECKMAP takes the 1D integrated light spectrum of a galaxy as input and computes its stellar population properties and kinematics. The stellar content is inferred by finding the linear combination of model spectra representing single stellar populations (SSPs) that best describes the observed spectrum. The weights of these model spectra give the age distribution of the stellar populations in the galaxy or, in other words, its SFH.
    
    The most important input parameters required for STECKMAP are the wavelength and age range, number of age bins, and the stellar population spectral library. For all galaxies we took the age range from 0.5 to 13.6\,Gyr, spanned by 30 bins, and a wavelength coverage of $4050$--$5500$\,\AA. We chose to use the \texttt{MILES} (Medium resolution INT\footnote{Isaac Newton Telescope} Library of Empirical Spectra) stellar population models, which are based on an extensive empirical stellar library with flux-calibrated spectra of 985 stars over a large range of stellar parameters \citep{sanchez2006}. It covers a large spectral range ($3525$--$7500$\,\AA) with a spectral resolution of $2.3$\,\AA\ (FWHM), but more importantly the stellar library is larger and more homogeneous in stellar parameter coverage than the libraries underlying the other models, and for these reasons \texttt{MILES} has become the community standard for nearby galaxy analysis.

    The stellar content outputs of STECKMAP are given as a function of age. The basic quantity from which all other stellar content outputs are determined is the stellar age distribution (SAD), which represents the normalized contribution of flux in each component to the observed spectrum. The stellar mass in each time bin is computed from the SAD and $M/L(\mathrm{age}_i , Z_i)$ ratio as a function of age and metallicity $Z$ in the given bin:
    \begin{equation}\label{eqn:mass}
        \mathrm{mass}_i = \frac{\mathrm{SAD}_i}{M/L(\mathrm{age}_i , Z_i)}.
    \end{equation}
    \noindent
    This stellar mass is the initial mass a given component has at the time of its birth because STECKMAP does not consider any mass loss; it only accounts for the dimming of the population and not for the recycling of or decrease in the stellar mass. The normalization of the stellar masses is arbitrary as the SAD is normalized to $\Sigma_i \mathrm{SAD}_i = 1$. Hence, the SFRs obtained from STECKMAP are not absolute but rather relative SFRs. They are computed from the fraction of the total mass of stars formed by a given time, not the fraction of the final stellar mass (which is less than the total mass of stars formed due to stellar mass loss) formed by that time.
    
    We obtain the (relative) SFR from the stellar mass ($\mathrm{mass}_i$) in each bin by dividing by the duration of age ($\Delta t_i$) in each bin (the extent of age in each bin is computed between midpoints of two adjacent bins: $\Delta t_i = t_{i+1/2} - t_{i-1/2}$): 
    \begin{equation}\label{eqn:sfr}
        \mathrm{SFR}_i = \frac{\mathrm{SAD}_i}{\Delta t_i}.
    \end{equation}
    \noindent
    
    \begin{figure}
    \centering
        \includegraphics[width=\columnwidth]{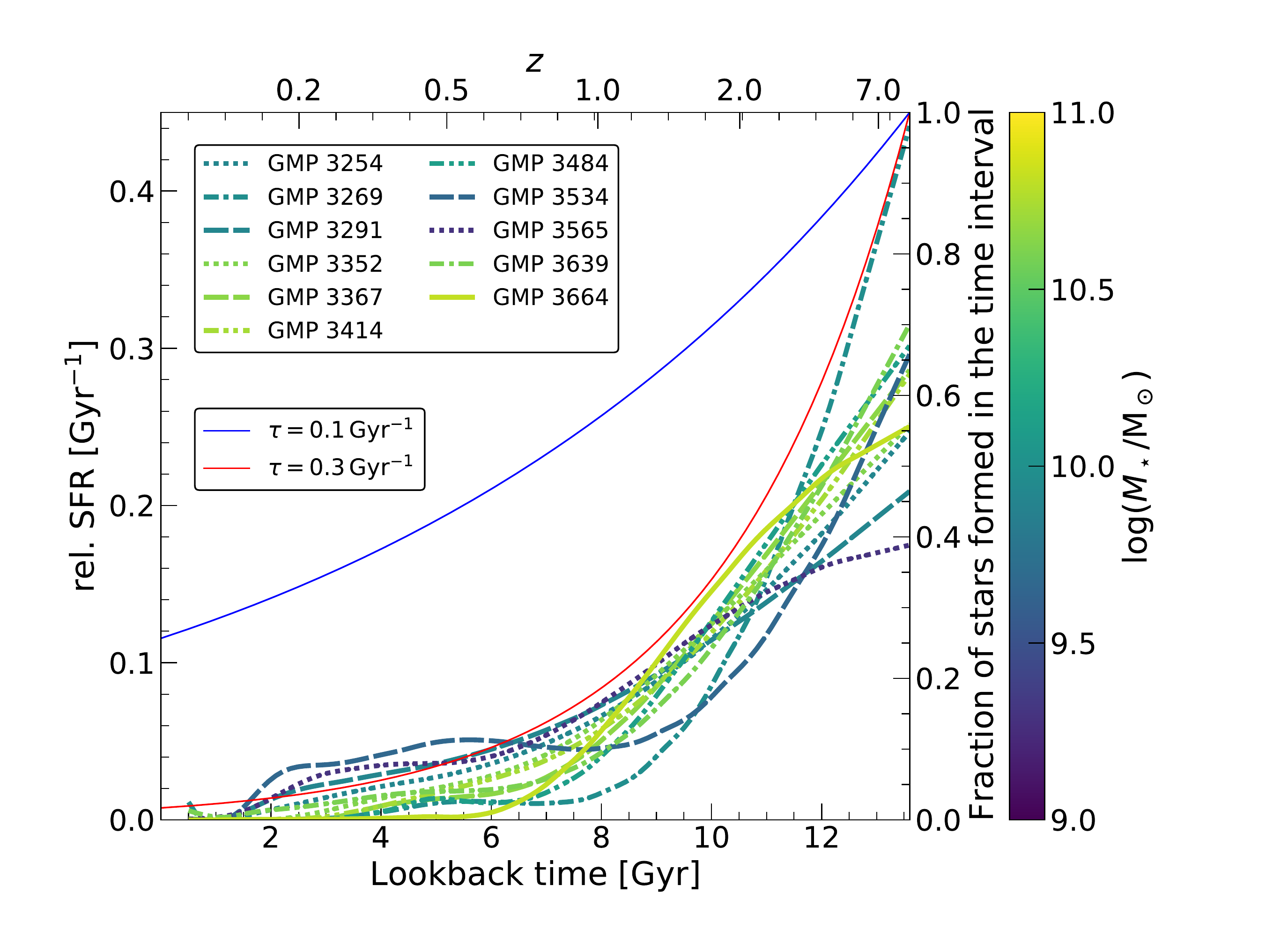}
    \caption{Star formation histories of galaxies in our sample determined using STECKMAP, using the \texttt{MILES} SSP model. The SFRs returned by STECKMAP are relative, not absolute, so we show them in units of $\textrm{Gyr}^{-1}$. The curves are color-coded based on the satellite stellar masses. STECKMAP does not cover lookback times $< 0.5$\,Gyr. The red and blue lines show the exponentially decaying form ($\mathrm{SFR} \propto \mathrm{exp}(\tau t)$) of SFHs of red (early-type) and blue (late-type) galaxies respectively \citep{Tojeiro2013}.}
    \label{fig:sfr_steckmap}
    \end{figure}
    
    The relative SFHs in units of $\mathrm{Gyr}^{-1}$ obtained from STECKMAP for our sample of galaxies are shown in Fig.~\ref{fig:sfr_steckmap}. They are color-coded by their stellar mass. As a visual guide, we also show two exponentially decaying rates $\mathrm{SFR} \propto \mathrm{exp}(\tau t)$, with decay timescales representative of early-type (red line, $\tau=0.3\,\mathrm{Gyr}^{-1}$) and late-type (blue line, $\tau=0.1\,\mathrm{Gyr}^{-1}$) galaxies in low-density environments \citep{Tojeiro2013}. At older ages the SFRs of massive galaxies are higher than those of their lower mass counterparts; this trend reverses at younger ages. We interpret this trend further below.
    
    All of the 11 galaxies in our sample were quenched $\gtrsim 2$\,Gyr ago and show no signs of recent star formation \citep{Trager_2008}. The slight upturn in SFR (see Fig.~\ref{fig:sfr_steckmap}) for some galaxies at $z \sim 0$ is likely to be due to the influence of a subtle effect in STECKMAP where young, metal-rich populations are substituted for old, metal-poor populations due to the age--metallicity degeneracy and missing hot-star populations in the underlying stellar population models \citep[see also][]{trager2005}.
    
    We computed stellar masses for our sample galaxies by multiplying their $B$-band luminosities by the color-dependent $M/L$ ratios of \citet{Bell_2003}, using the  $B-R$ colors measured by \citet{eisenhardt2007}. We assumed a distance modulus of $34.94$ to Coma and a solar absolute $B$-band magnitude of $5.51$.

 \subsection{Orbit libraries from cosmological simulations}\label{sec:orbits}
    We inferred orbital parameters for our sample of Coma galaxies by extracting orbital parameter distributions drawn from an N-body simulation, following the general approach of \citet{Oman2021}. We used the same initial conditions as the Level--0 simulation described in  \citet{wang2019}. The simulation assumes a $\Lambda$CDM cosmology with Planck 2014 \citep{planck2014} parameters: mean matter density $\Omega_m = 0.307$, cosmological constant $\Omega_{\Lambda} = 0.693$, and Hubble parameter at the current epoch $H_0 = 66.7 \, \mathrm{km \, s^{-1}\, Mpc^{-1}}$. We evolved the initial conditions up to a scale factor $a=2$ ($z = -0.5$, $\approx 10$\,Gyr into the future) to predict the time of the first pericentric passage in cases where it has not yet occurred. The simulation has a cube length of $738$\,Mpc, a particle mass of $1.55 \times 10^9\,\mathrm{M}_\odot$, and a maximum time between consecutive outputs of $390$\,Myr. The simulation was run using the \texttt{GADGET4} code \citep{Springel_2020}. The simulation was processed by first finding the halos using the \texttt{ROCKSTAR} \citep{rockstar} halo finder and then linking the halos in merger trees using the \texttt{CONSISTENT--TREES} \citep{mergertree} code. 
    
    We selected a sample of ``Coma-like'' host systems on the basis of their halo mass (details of this selection are given in the next subsection). We label as satellites the halos within a 3D aperture of $2.5$ times the virial radius $r_\mathrm{vir}$ at $z=0$ around the Coma-like hosts. The primary progenitors of the hosts and satellites were traced both forward and backward in time using the merger trees in order to obtain the orbital histories of the satellites. We tabulated the ``observable coordinates'' of the satellites at $z=0$, namely: the projected offset from cluster center normalized by the cluster virial radius ($R/r_\mathrm{vir}$) and the line-of-sight velocity offset from cluster center normalized by the velocity dispersion of dark matter particles in the host halo ($V/\sigma_\mathrm{3D}$). We arbitrarily chose the third spatial axis ($z$--axis) of the simulation as that parallel to the line of sight. Here, $R$ and $V$ are given in terms of simulated system coordinates by
    
    \begin{equation}\label{eqn:rR}
       \left(\frac{R}{r_\mathrm{vir}}\right) = \frac{\sqrt{(r_\mathrm{host,x}-r_\mathrm{sat,x})^2 + (r_\mathrm{host,y}-r_\mathrm{sat,y})^2}}{r_\mathrm{vir}}
    \end{equation}
    and
    \begin{equation}\label{eqn:vV}
       \left(\frac{V}{\sigma_\mathrm{3D}}\right) = \frac{|v_\mathrm{host,z} - v_\mathrm{sat,z}| + H(r_\mathrm{host,z} - r_\mathrm{sat,z})}{\sigma_\mathrm{3D}},
    \end{equation}
    
    \noindent 
    where $r$ and $v$ are the coordinate and velocity of the satellites in the simulated system coordinates, the $x$, $y$ and $z$ subscripts denote their components along orthogonal axes, $H$ is the Hubble parameter, $r_\mathrm{vir}$ is the virial radius used to normalize $R$ and $\sigma_\mathrm{3D}$ is the velocity dispersion used to normalize $V$. 
    
    For each satellite, we also tabulated the maximum virial mass at any past time ($M_\mathrm{max}$), the time of first infall through $2.5 r_\mathrm{vir}$ ($t_\mathrm{inf}$), and the time of first pericenter ($t_\mathrm{peri}$). We also selected ``interlopers'' in the foreground and background of each Coma-like halo. These fall within $2.5 r_\mathrm{vir}$ in projection at $z=0$, but actually are outside $2.5 r_\mathrm{vir}$ in 3D. We treated these similarly to the selected satellites but did not trace their orbits, instead using them only to determine an ``interloper probability'' for each of our observed galaxies.

\subsection{Orbital parameter estimation for observed satellites}\label{sec:obsorbits}
    We then proceeded to select possible orbits for each observed galaxy from the distribution of simulated orbits. We selected these by performing cuts on the library of simulation orbits based on the observed properties of the Coma satellites.

    We begin by determining the virial mass and velocity dispersion of the Coma cluster, needed to normalize the PPS coordinates of the satellites. We assume a virial radius of $r_\mathrm{vir,Coma}=2.9\,h_{70}^{-1}\mathrm{\,Mpc}$ \citep{Lokas_2003}, which assumes the same definition of the virial radius as that used in our orbit libraries. The corresponding virial mass is $M_\mathrm{host,Coma}=1.53 \times 10^{15}\,\mathrm{M}_\odot$, computed from the virial radius as $M_{\mathrm{vir}} = \frac{4\pi}{3} \Delta_{\mathrm{vir}}(z)\,\Omega_{\mathrm{m}}(z)\,\rho_{\mathrm{crit}}(z)\,r_{\mathrm{vir}}^3$, where $\Delta_{\mathrm{vir}}$ is the virial overdensity in units of the mean matter density $\Omega_{\mathrm{m}}\rho_{\mathrm{crit}}$ and $\rho_{\mathrm{crit}}={3H^2}/(8\pi G)$. We use the redshift of GMP~3329 ($z=0.023$), which is taken to be at the location of the X\nobreakdash-ray center of the Coma cluster\footnote{The X-ray position given by \citet{Ebeling1998} is just over $3^\prime$ away from GMP~3329, or about $3\sigma$. However, given the structural complexity of the Coma cluster, taking GMP~3329 as the cluster center simplifies the analysis here without substantially affecting the final results.}, as the cluster's systemic redshift. We then compute the velocity dispersion of the Coma cluster following \citet[][see also \citealp{Bryan1998} for the redshift dependence]{biviano2006}, noting that this velocity dispersion corresponds closely to the dark matter particle velocity dispersion used in the orbit libraries \citep[within about 10\%,][]{Munari_2013}:
    \begin{equation}\label{eqn:sig1d}
        \frac{\sigma_{1\mathrm{D}}}{\mathrm{km}\,\mathrm{s}^{-1}} = \frac{0.0165}{\sqrt{3}}\left(\frac{M_\mathrm{vir}}{\mathrm{M}_\odot}\right)^{\frac{1}{3}}\left(\frac{\Delta_\mathrm{vir}(z)}{\Delta_\mathrm{vir}(0)}\right)^{\frac{1}{6}}(1+z)^{\frac{1}{2}}.
    \end{equation}
    \noindent
    Finally we compute $\sigma_\mathrm{3D}$ from $\sigma_\mathrm{1D}$ assuming an isotropic velocity dispersion, $\sigma_\mathrm{3D} = \sqrt{3} \sigma_\mathrm{1D}$, using the computed value of $\sigma_\mathrm{1D}=1008 \,\mathrm{km\,s^{-1}}$, which gives $\sigma_\mathrm{3D} = 1745 \,\mathrm{km\,s^{-1}}$. 
    
    We turn our attention next to estimating the virial masses of the satellite galaxies, using the stellar-to-halo mass relation (SHMR) of \citet[][Eqn.~21]{Behroozi_2010}:
    \begin{equation}\label{eqn:sm-hm}
        \log(M_\mathrm{h}(M_\star)) = \log(M_1) + \beta \log\left(\frac{M_\star}{M_{\star,0}}\right) + \frac{\left(\frac{M_\star}{M_{\star,0}}\right)^\delta}{1+\left(\frac{M_\star}{M_{\star,0}}\right)^{-\gamma}} - \frac{1}{2},
    \end{equation}
    \noindent
    where $M_\mathrm{h}(M_\star)$ is the halo mass for which the average stellar mass is $M_\star$, $M_1$ is the characteristic halo mass ($\log(M_1) = 12.35$ at $z=0$), $M_{\star,0}$ is the characteristic stellar mass ($\log(M_{\star,0}) = 10.72$ at $z=0$), and the constants are power-law coefficients for the relation (at $z=0$: $\beta = 0.44$, $\delta = 0.57$ and $\gamma = 1.56$).
    
 \begin{table*}
    \centering
        \caption[]{Orbital properties of Coma cluster galaxies.}
        \label{tab:orb_prop}
        \begin{tabular}{ccccccc}
            \hline
            \noalign{\smallskip}
            GMP~& $\log(M_\mathrm{h}/\mathrm{M}_\odot)$ & $z_\mathrm{g}$ & $\Delta \theta$ ($^{\circ}$) & $R/r_\mathrm{vir}$ & $V/\sigma_\mathrm{3D}$ & Interloper probability (\%)\\
            \noalign{\smallskip}
            \hline
            3254 & 11.51 & 0.0251 & 0.019 & 0.011 & 0.180 & 1.9 \\ 
            3269 & 11.54 & 0.0267 & 0.015 & 0.009 & 0.432 & 0.0 \\ 
            3291 & 11.51 & 0.0226 & 0.030 & 0.017 & 0.202 & 0.0 \\ 
            3352 & 12.16 & 0.0200 & 0.014 & 0.008 & 0.009 & 3.1 \\ 
            3367 & 12.21 & 0.0240 & 0.027 & 0.016 & 0.702 & 3.2 \\ 
            3414 & 12.37 & 0.0225 & 0.022 & 0.013 & 0.226 & 0.0 \\ 
            3484 & 11.62 & 0.0159 & 0.041 & 0.023 & 0.539 & 0.0 \\ 
            3534 & 11.39 & 0.0224 & 0.055 & 0.032 & 0.585 & 0.8 \\ 
            3565 & 11.22 & 0.0239 & 0.061 & 0.035 & 0.015 & 3.1 \\
            3639 & 12.13 & 0.0193 & 0.077 & 0.045 & 1.209 & 2.2 \\ 
            3664 & 12.56 & 0.0239 & 0.086 & 0.050 & 0.213 & 0.0 \\
            \noalign{\smallskip}
            \hline
        \end{tabular}
    \tablefoot{The halo mass of each galaxy in solar mass units ($\log M_\mathrm{h}$), its redshift ($z_\mathrm{g}$), angular separation from host center in degrees ($\Delta \theta$), normalized projected distance from host center ($R/r_\mathrm{vir}$), and normalized projected velocity ($V/\sigma_\mathrm{3D}$) are tabulated.}
    \end{table*}

    
    \begin{figure}
    \centering
        \includegraphics[width=\columnwidth]{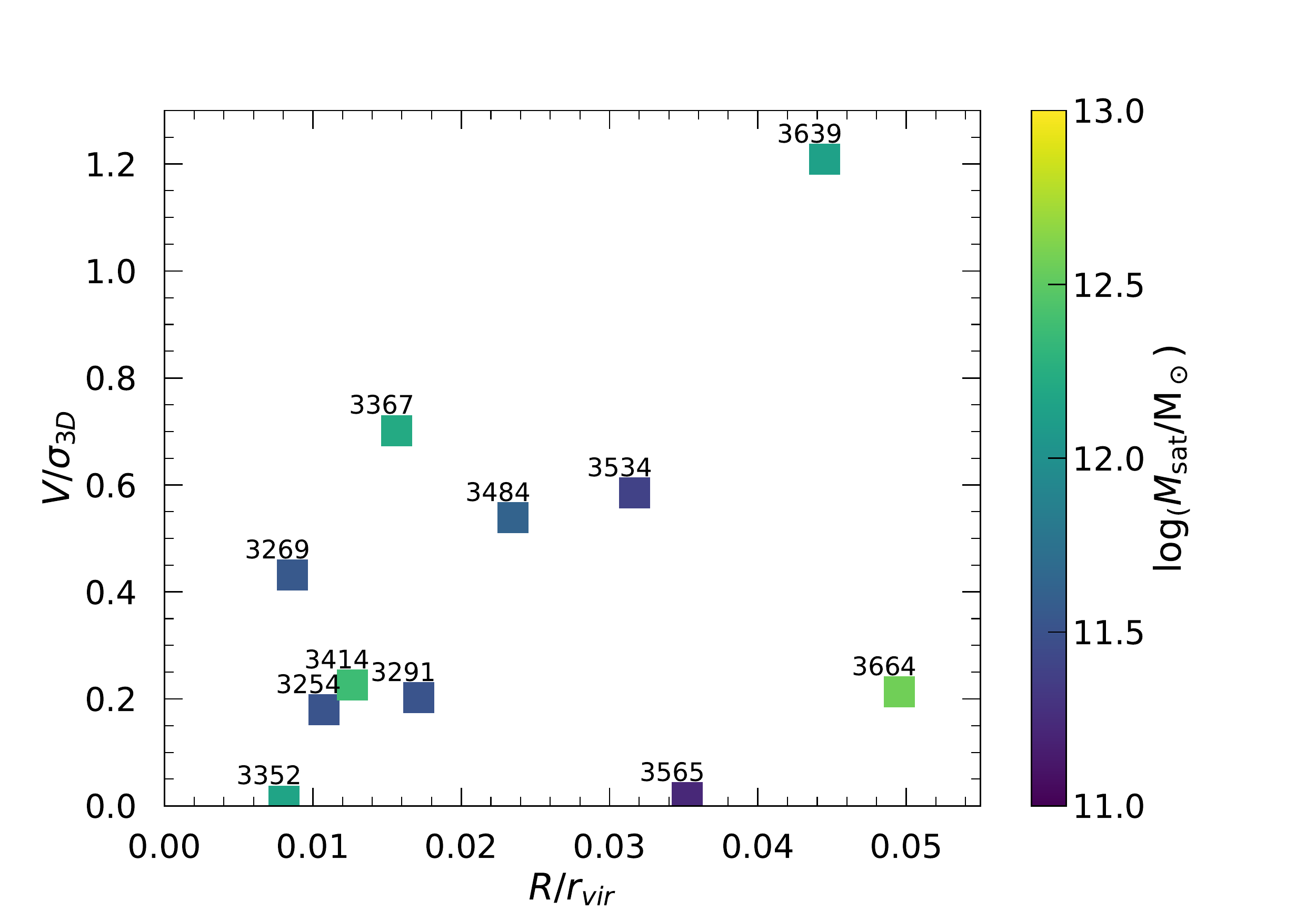}
    \caption[Position of galaxies in projected phase space]{Galaxies used in this study shown in projected phase space. The galaxies are colored by $\log(M_{sat}/\mathrm{M_\odot})$. The points are labeled with their GMP identifiers.}
    \label{fig:pps}
    \end{figure}
    
    The virial masses estimated in this way do not account for tidal stripping of the satellites' dark matter halos: We interpret these as the virial masses that the satellites had before entering the cluster, and will match them against the maximum mass of the satellite halos in the simulations. This approximation is imperfect for two reasons. First, we use the $z=0$ SHMR. We should in principle use the SHMR at the redshift corresponding to the infall time of the satellites, but to estimate this infall time requires us to know their virial masses. Second, we neglect any stellar mass growth or loss between infall and the present.  Although we do have SFH information, a self-consistent accounting for the stellar mass growth again requires knowledge of the orbital history, which we are in the process of attempting to estimate. For stellar mass loss, we assume that any tidal stripping of stars has been moderate \citep[see][]{Smith_2016}: Once satellites are heavily stripped of stars, they are very quickly destroyed \citep{Bahe_2019}. We will explore the extent to which the systematic offsets introduced by this approximation may affect our results in Sect.~\ref{sec:disc-shmr}.
    
    The PPS coordinates of each satellite, $R/r_\mathrm{vir}$ and $V/\sigma_\mathrm{3D}$, are computed from their measured sky coordinates and redshifts as
    \begin{equation}\label{eqn:rcut}
        \frac{R}{r_\mathrm{vir}} = \frac{d_\mathrm{A} \Delta \theta}{r_\mathrm{vir}}
        \end{equation}
        and
        \begin{equation}
        \frac{V}{\sigma_\mathrm{3D}} = \frac{c|z_\mathrm{g} - z_\mathrm{c}|}{(1+z_\mathrm{c})\sigma_\mathrm{3D}},
    \end{equation}
    \noindent where $d_\mathrm{A} = 99\,\mathrm{Mpc}$ is the angular diameter distance of the Coma cluster and $\Delta \theta$ is the angular separation of the satellites from the Coma center (taken to be the angular distance from GMP~3329, as above), and where $c$ is the speed of light, $z_\mathrm{g}$ is the redshift of the satellite, $z_\mathrm{c}$ is the redshift of the Coma cluster and $\sigma_\mathrm{3D}$ is the 3D velocity dispersion of the Coma cluster (Eq.~\ref{eqn:sig1d}). The measured and computed properties for satellites of Coma are listed in Table~\ref{tab:orb_prop} for reference. Fig.~\ref{fig:pps} shows the Coma galaxies in projected phase space. We note that the spread in the radii is very small. The orbits in our N-body libraries are not significantly different between satellites so close together in radius, so differences between the inferred orbits are driven primarily by the satellite velocity offsets ($V/\sigma_\mathrm{3D}$), and their halo masses. 
    
    We select a distribution of likely orbits for each satellite by selecting orbits from the simulation that have the same PPS coordinates and host and satellite virial masses, within the following tolerances. For $M_\mathrm{host}$ and $M_\mathrm{max}$, the upper and lower limits of the selections are set $0.5\,\mathrm{dex}$ above and below the virial mass of the Coma cluster and the satellite, respectively. The limits for the selections around the PPS coordinates $R/r_\mathrm{vir}$ and $V/\sigma_\mathrm{3D}$ are set to $\pm 0.05$ of their measured values. The cuts resulted in many orbits for satellites that are similar (in terms of mass and current projected coordinates) to the observed galaxies. We then make the reasonable assumption that the selected orbits represent an approximation of the probability distribution of possible orbits for a given satellite. The parameters used in this study are listed in Table~\ref{tab:orb_prop}. 
    
    The Coma cluster has the interesting property of having two similarly bright BCGs, GMP~3329 (NGC~4874) and NGC~4889. We find a reasonable number of Coma like clusters (458) in the simulation that could be split into two roughly even groups based on whether they have a very massive satellite near the center. Of these clusters, 126 have a satellite with mass at infall $> 1\times10^{14}\,\mathrm{M}_\odot$, which might be reasonable matches for the halo mass of NGC~4889 at infall, though this mass is difficult to constrain observationally. However, as the number of matches for each satellite in our sample is rather small ($\lesssim 100$), splitting the cluster sample would strain our ability to construct orbital parameter PDFs for the sample of satellites. In light of these consideration, we do not explore the topic of Coma's substructure further in this study; analysis of a larger-volume simulation where a large number of double-BCG systems could be analyzed could be pursued in future work. Another possible complication is that the halo mass function is rather steep, and we have used a symmetrical selection criterion of $\pm 0.5\,\mathrm{dex}$ from the target values for cluster and satellite halos; however, we shown in Sect.~\ref{sec:systematic} below that this steepness makes no appreciable impact on our results.
    
    We make analogous selections from the library of interloper halos. By comparing the number of interlopers selected to the number of satellite orbits, we obtain an estimate of the probability that the satellite is in fact an interloper. These interloper probabilities are included self-consistently throughout the rest of the analysis described below, but they have little influence because they are very small (see Table~\ref{tab:orb_prop}). We therefore do not discuss this point further.

\section{Results}\label{sec:results}
\subsection{PDFs of orbital time and SFR comparison}\label{subsec:pdf_sfr}

    \begin{figure*}
    \centering
        \includegraphics[width=0.49\textwidth]{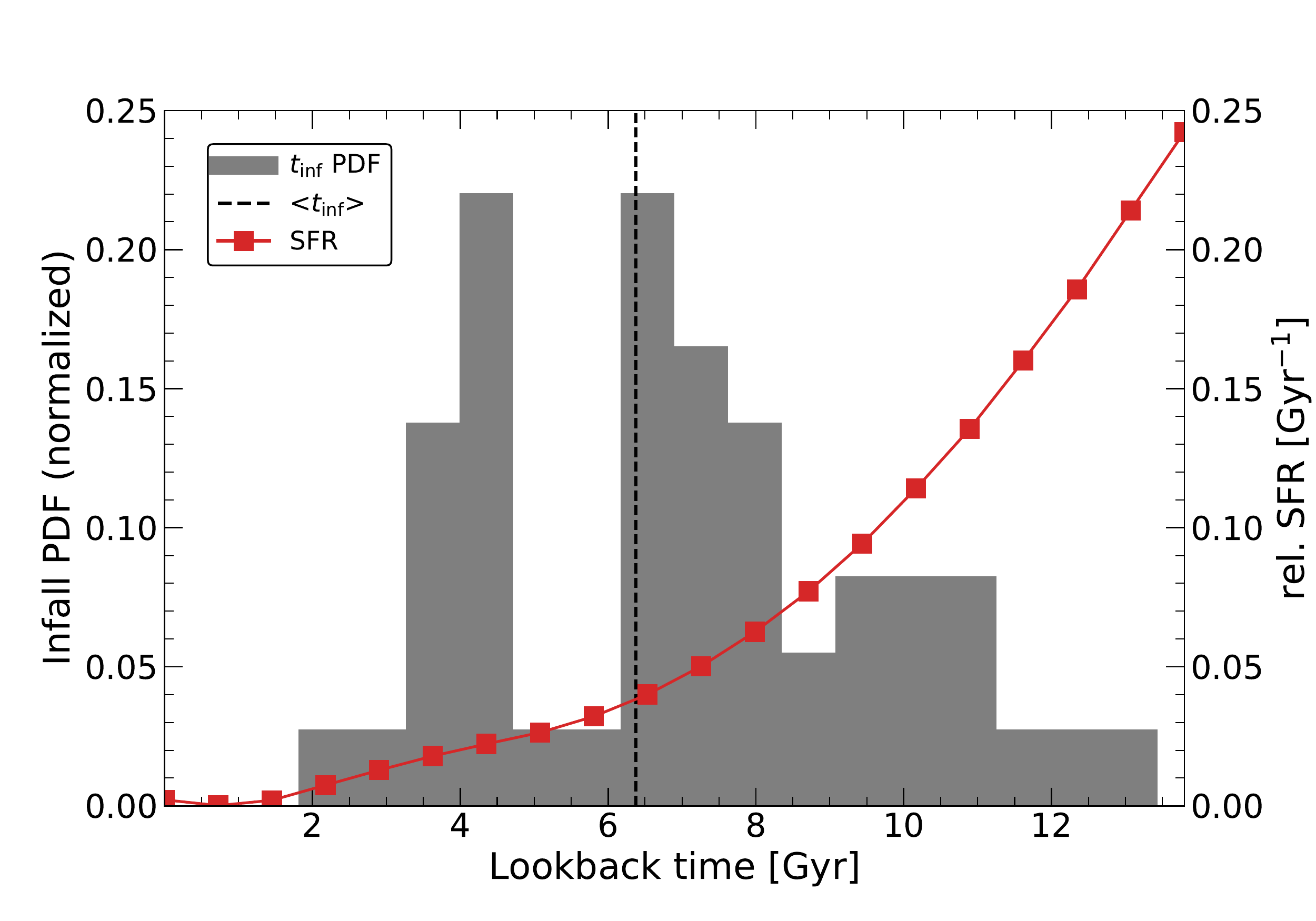}
        \includegraphics[width=0.49\textwidth]{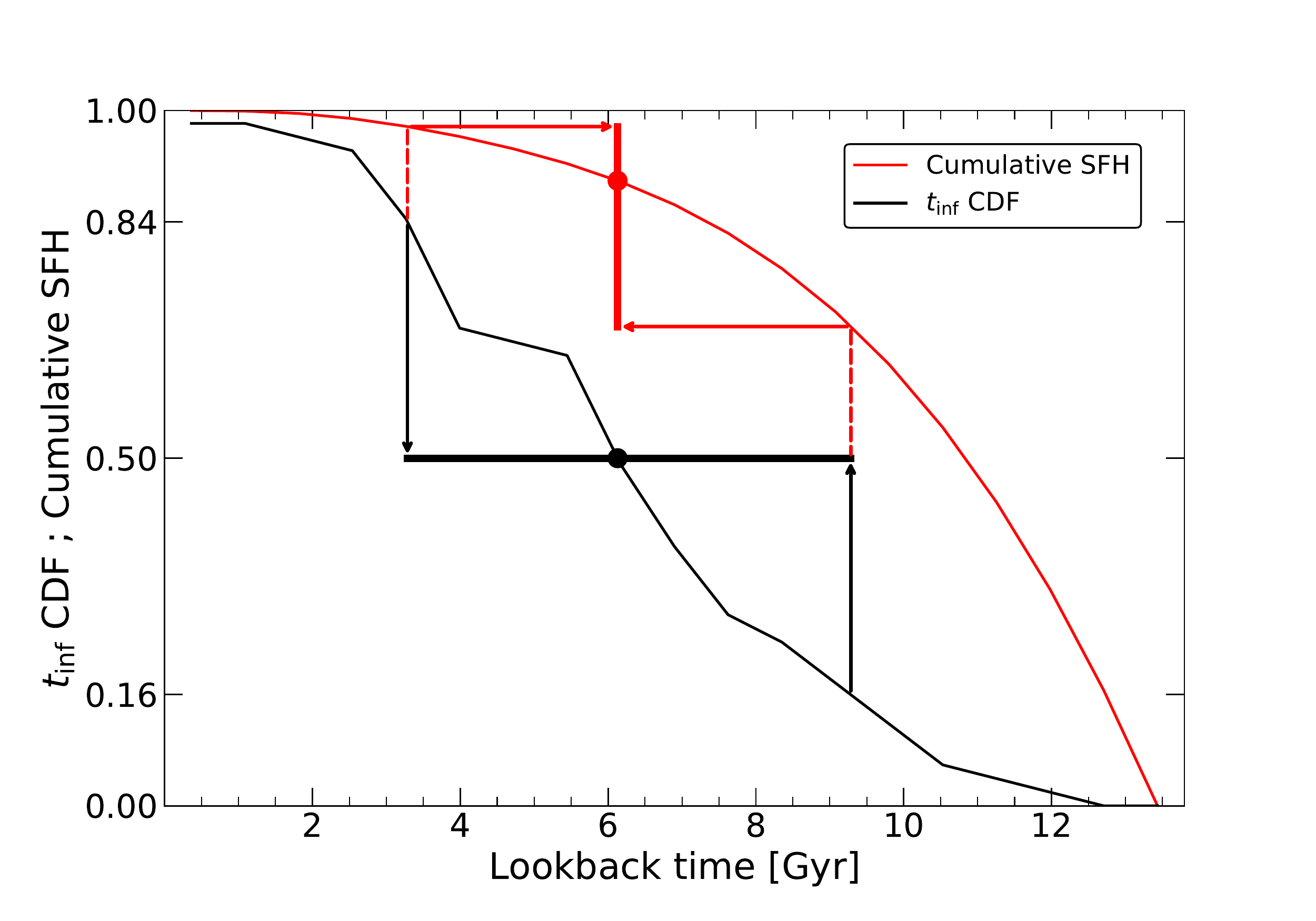}
    \caption[]{Connecting orbital history with SFH. 
    \emph{Left panel}:~Probability distribution of infall time of GMP~3254 is shown in the gray histogram (left axis labels) and its SFH (right axis labels) is shown as the red points. The median value of its infall time probability distribution is shown as a dotted black vertical line. The SFR has dropped significantly by $\langle t_\mathrm{inf}\rangle$.
    \emph{Right panel}:~Cumulative SFH -- the fractional stellar mass accumulation over time -- is shown (red line) along with the cumulative probability distribution of the infall time (black line) for GMP~3254. The median infall time (black filled circle) is marked with its 68\% confidence interval (horizontal black line) at the $50^\mathrm{th}$ and $16^\mathrm{th}$--$84^\mathrm{th}$ percentiles, respectively. The corresponding accumulated stellar mass at the median infall time (red filled circle) and the its associated 68\% confidence interval (vertical red line) are also shown. The arrows (black and red) indicate the translation of $16^\mathrm{th}$ and $84^\mathrm{th}$ percentile points on the cumulative distribution curve to boundary of the confidence interval lines. GMP~3254 ($\log_{10}M_\star/\mathrm{M}_\odot=9.92$) has formed around 90\% of its stellar mass at the median infall time. Fig.~\ref{fig:final} (left panel) shows the accumulated stellar mass at the (median) infall and pericenter times for all galaxies in this study.}
    \label{fig:cdf-pdf}
    \end{figure*}
    
    We illustrate our approach to compare the SFHs and orbital histories of our sample of Coma satellites in Fig.~\ref{fig:cdf-pdf}, using GMP~3254 as a representative example. In the left panel, we show the differential probability distribution for the infall time of this satellite (gray histogram), as determined from the N-body orbit library, and the differential SFH (red line). 
    
    The infall time probability distribution is broad and has two peaks: The younger peak corresponds to the possibility that GMP~3254 is now near its first pericentric passage in Coma, while the older peak corresponds to the possibility that it is near its second pericentric passage (the tail to even older ages corresponds to multiple orbits having been completed). The separation of the two peaks reflects (twice) the crossing time of the Coma cluster $t_\mathrm{cross}={r_\mathrm{vir}}/{\sigma_\mathrm{1D}}\sim 2.5$\,Gyr, keeping in mind that the crossing time would be somewhat shorter for earlier infall times. The SFH is monotonically declining toward the present time, and has no clear or sharp features. This makes it difficult to immediately see any obvious link between the SFH and orbital history.
    
    \begin{figure*}
        \includegraphics[width=\textwidth]{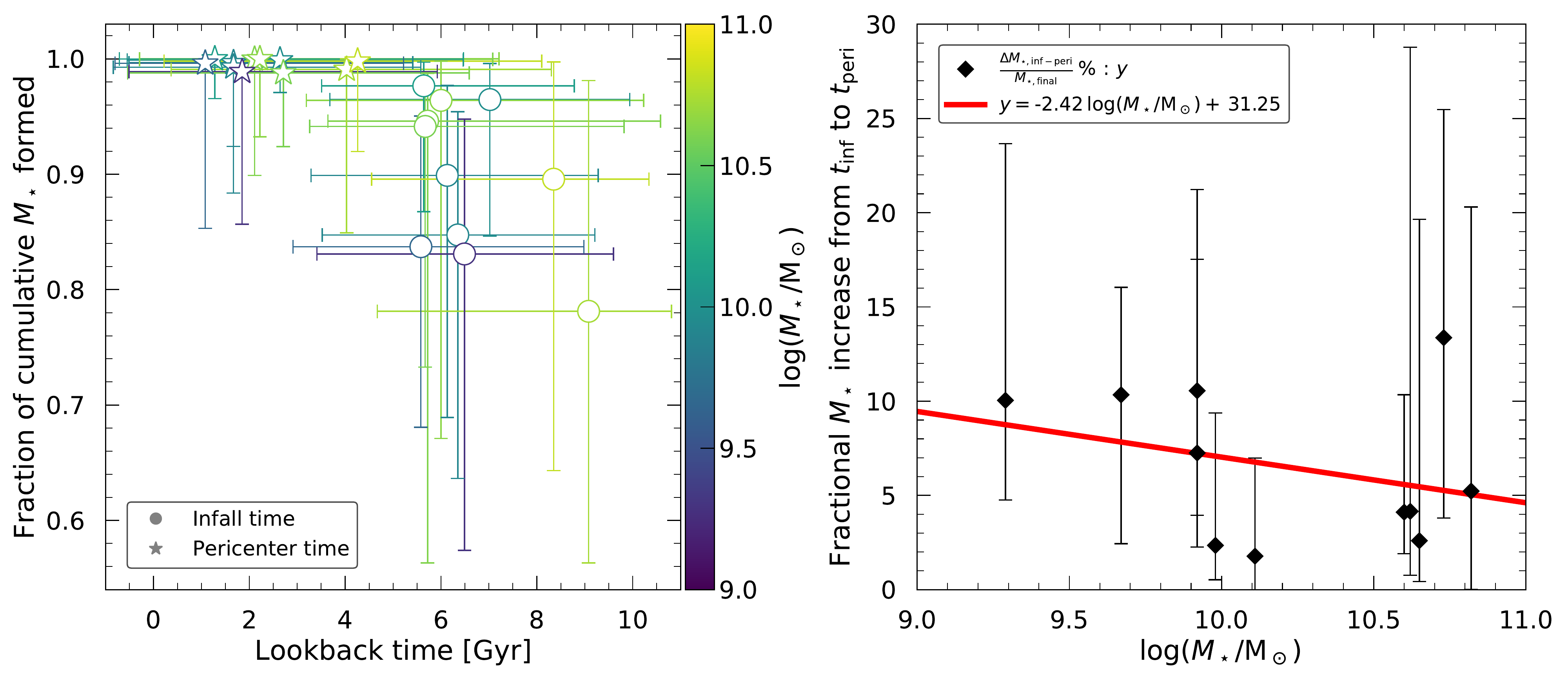}
    \caption[]{The stellar mass formed in the Coma cluster environment as a function of orbital phase. 
    \emph{Left panel}:~Fraction of cumulative stellar mass for all the galaxies in this study at their expected infall (circles) and pericenter (stars) times. The galaxies are color-coded based on their present-day stellar masses. The error bars represent $68\%$ confidence intervals on the expectation values. Higher stellar mass galaxies have formed a higher percentage of stellar mass at infall in comparison to the lower stellar mass galaxies. All galaxies have likely formed more than $95\%$ of their stellar mass by the time of their first pericentric passage in Coma.
    \emph{Right panel}:~Fractional increases in stellar mass between infall and pericenter time (black filled diamonds with $68\%$ confidence intervals) are shown for all galaxies. The medians and confidence intervals are determined from the distribution of time differences and corresponding assembled mass differences computed for each individual matching N-body orbit (i.e., we do not plot the difference between the marginalized distributions shown in the left panel). The linear regression through the data (magenta line) hints that lower stellar mass galaxies may have formed relatively more stellar mass as compared to their higher stellar mass counterparts.}
    \label{fig:final}
    \end{figure*}
     
    We also computed the differential probability distribution for pericenter time and compared it with the relative SFR, following the same method illustrated in the left panel of Fig.~\ref{fig:cdf-pdf} for the infall time. However, it is difficult to see any clear trends because the probability distributions are broad and the inferred SFHs are smooth.

    \begin{table*}
    \centering
        \caption[]{Expected infall and pericenter times for Coma cluster satellites and the accumulated stellar mass at those times.}
        \label{tab:inf-peri}
        \begin{tabular}{l r r r r r}
            \hline
            \noalign{\smallskip}
            GMP~& $\log(M_\star/\mathrm{M_\odot})$ & $t_\mathrm{inf} \, \mathrm{(Gyr)}$ & $t_\mathrm{peri} \, \mathrm{(Gyr)}$ & $M_\star \, \mathrm{(\%)} \, \mathrm{at} \, t_\mathrm{inf}$ & $M_\star \, \mathrm{(\%)} \, \mathrm{at} \, t_\mathrm{peri}$ \\
            \noalign{\smallskip}
            \hline
            3254 & 9.92 & $6.1^{+3.2}_{-2.8}$ & $1.7^{+3.7}_{-2.2}$ & $89.90^{+7.79}_{-20.98}$ & $99.66^{+0.34\phantom{0}}_{-7.24\phantom{0}}$ \\ \\
            3269 & 9.98 & $7.0^{+2.9}_{-3.3}$ & $2.6^{+3.8}_{-3.2}$ & $96.47^{+3.12}_{-11.82}$ & $99.90^{+0.10\phantom{0}}_{-2.82\phantom{0}}$ \\ \\
            3291 & 9.92 & $6.4^{+2.9}_{-2.8}$ & $1.7^{+3.9}_{-2.5}$ & $84.74^{+10.67}_{-21.10}$ & $99.28^{+0.72}_{-10.92}$ \\ \\
            3352 & 10.62 & $5.7^{+4.9}_{-2.1}$ & $2.1^{+5.0}_{-2.4}$ & $94.60^{+4.14}_{-38.29}$ & $99.96^{+0.04}_{-10.07}$ \\ \\
            3367 & 10.65 & $6.0^{+4.2}_{-2.8}$ & $2.2^{+5.0}_{-2.5}$ & $96.40^{+3.39}_{-29.30}$ & $100.00^{+0.00\phantom{0}}_{-6.76\phantom{0}}$ \\ \\
            3414 & 10.73 & $9.1^{+1.7}_{-4.4}$ & $4.0^{+4.3}_{-3.7}$ & $78.12^{+19.99}_{-21.81}$ & $99.08^{+0.92}_{-14.16}$ \\ \\
            3484 & 10.11 & $5.6^{+3.1}_{-2.1}$ & $1.3^{+5.2}_{-2.0}$ & $97.66^{+2.05}_{-10.91}$ & $99.99^{+0.01\phantom{0}}_{-3.44\phantom{0}}$ \\ \\
            3534 & 9.67 & $5.6^{+3.4}_{-2.7}$ & $1.1^{+4.1}_{-1.9}$ & $83.72^{+11.33}_{-15.65}$ & $99.62^{+0.38}_{-14.31}$ \\ \\
            3565 & 9.29 & $6.5^{+3.1}_{-3.1}$ & $1.9^{+4.1}_{-2.4}$ & $83.09^{+11.69}_{-25.69}$ & $98.89^{+1.11}_{-13.23}$ \\ \\
            3639 & 10.60 & $5.7^{+4.1}_{-2.4}$ & $2.7^{+3.9}_{-3.3}$ & $94.14^{+3.92}_{-20.85}$ & $98.78^{+1.22\phantom{0}}_{-6.38\phantom{0}}$ \\ \\
            3664 & 10.82 & $8.4^{+2.0}_{-3.8}$ & $4.8^{+3.3}_{-4.6}$ & $89.57^{+10.17}_{-25.26}$ & $99.70^{+0.31\phantom{0}}_{-7.73\phantom{0}}$ \\
            \noalign{\smallskip}
            \hline
        \end{tabular}
    \tablefoot{The expected value of the infall and pericenter times are listed with their uncertainties in the central $68\%$ confidence interval. The accumulated stellar mass at the expected infall and pericenter times and at the corresponding boundaries of the uncertainty intervals are also given.}
    \end{table*}
     
    To bring out subtler trends, we next  quantify the probable fraction of GMP~3254's stellar mass assembled by the time it fell into Coma. This is illustrated in the right panel of Fig.~\ref{fig:cdf-pdf}. The expected infall time is the time where its cumulative distribution (black curve) crosses $0.5$, and the $68$\% confidence interval on this expectation (thick black bar) is bounded by the locations where the cumulative distribution crosses $0.16$ and $0.84$. The cumulative fraction of the stellar mass formed, according to the STECKMAP SFH, is shown with the red curve. At the expected infall time ($\sim 6$\,Gyr), GMP~3254 had formed $\sim 90$\% of its final stellar mass, while at the boundaries of the $68$\% confidence interval on the infall time, it had formed $\sim 69$\% and $\sim 98$\% of its final stellar mass -- we interpret this range as an approximate $68$\% confidence interval on the stellar mass assembled at the time of infall. We repeat the same measurement replacing the infall time probability distribution with the probability distribution for the time of the first pericentric passage to estimate the fraction of stellar mass assembled by that time, and its associated confidence interval, and repeat the same measurement for each galaxy in our sample. The resulting values are listed in Table~\ref{tab:inf-peri}.

    The left panel of Fig.~\ref{fig:final} shows the fraction of cumulative stellar mass formed at the expected infall (circles) and pericenter (stars) times for all galaxies in this study. The $68\%$ confidence interval for the expected orbital times and the corresponding fractions of cumulative stellar mass are also shown. The points are colored according to the present-day stellar masses of the satellites. The satellites in our sample likely fell into the Coma cluster at lookback times of around $6$--$9$\, Gyr and had their first pericentric passages at lookback times of $1$--$5$\,Gyr, with the time between infall and pericenter ranging between $3$--$5$\,Gyr.
    
    We find tentative evidence for a trend in the fractional mass formed after infall as a function of (present-day) galaxy mass, as illustrated in the right panel of Fig.~\ref{fig:final}. More massive galaxies seem to have formed a larger fraction of their final stellar mass relative to the lower mass galaxies at their expected infall times. The galaxies with higher stellar mass ($\log(M_\star/\mathrm{M_\odot}) \gtrsim 10$) have typically formed more than $\sim 90\%$ of their stellar mass by their expected infall time, while the lower mass galaxies have mostly formed around $80$--$90\%$ of their stellar mass. However, GMP~3414 ($\log(M_\star/\mathrm{M}_\odot)=10.73$) and GMP~3664 ($\log(M_\star/\mathrm{M}_\odot)=10.82$), which are near the high-mass end of our sample, are outliers to this trend; they likely fell in $\sim 2$\,Gyr earlier than the other satellites. One plausible explanation for this could be that these high mass galaxies formed most of their stars early, so when these two fell in (early), they were at their peak of star formation and therefore somewhat more resilient to the cluster environment. The cluster environment was also perhaps a bit gentler earlier on, but this is probably not the main driver of this trend. We note that \citet{Wetzel_2013} have also highlighted that star formation continues after infall.
    
    
    All the sample galaxies likely formed more than $\sim 98\%$ of their stellar mass by the time of their first pericentric passage and thus had nearly completely quenched by that time. Some galaxies have a nonzero probability of currently being on their first infall, which is translated into error bars for the pericenter time dropping below zero in some cases (see left panel of Fig.~\ref{fig:final}). Despite the wide confidence intervals, our measurements hint that these galaxies likely formed a significant fraction ($\gtrsim 5\%$) of their stellar mass while in the high-density environment of the Coma cluster.

\section{Discussion}

\subsection{Quenching in the Coma cluster}

    Our results suggest that the galaxies we have studied in the Coma cluster are quenched near their first pericenter passage. Furthermore, the galaxies with higher stellar masses likely entered the cluster with a slightly larger fraction of stellar mass already formed; lower stellar mass galaxies formed roughly twice as much of their total stellar mass after infall. Using our main result shown in Fig.~\ref{fig:final} (left panel) and comparing them to previous studies, we now discuss possible quenching processes along the orbits of the satellites.
    
    \begin{table}
    \centering
        \caption[]{Quenching time after first pericentric passage.}
        \label{tab:tq}
        \begin{tabular}{ccc}
            \hline
            \noalign{\smallskip}
            GMP~& $t_\mathrm{q}$ after $t_\mathrm{peri}$\,[Gyr] & $t_\mathrm{q}$ after $t_\mathrm{peri}$\,[Gyr] \\
             & using lim. SSFR & using avg. SSFR \\
             & of $10^{-11}\,\mathrm{yr}^{-1}$ & of $3\times 10^{-10} \,\mathrm{yr}^{-1}$ \\
            \noalign{\smallskip}
            \hline
            3254 & 0.34 & 0.01 \\
            3269 & 0.10 & 0.00 \\
            3291 & 0.72 & 0.02 \\
            3352 & 0.04 & 0.00 \\
            3367 & 0.00 & 0.00 \\
            3414 & 0.92 & 0.03 \\
            3484 & 0.01 & 0.00 \\
            3534 & 0.38 & 0.01 \\
            3565 & 1.11 & 0.04 \\
            3639 & 1.22 & 0.04 \\
            3664 & 0.30 & 0.01 \\
            \noalign{\smallskip}
            \hline
        \end{tabular}
    \tablefoot{The time required after the first pericenter for all the galaxies in our study to form 100\% of their stellar mass assuming the limiting SSFR for active star formation of $10^{-11}\,\mathrm{yr}^{-1}$, and a typical rate for star forming galaxies at $z\sim 0.25$ of $\mathrm{SSFR}= 3\times 10^{-10} \,\mathrm{yr}^{-1}$ \citep{Madau_2014}. All the galaxies have completely quenched within a few\,Myr to $\sim 1$\,Gyr after the first pericenter using either of the two approaches.}
    \end{table}
    
    To guide our interpretation, we first ask, based on our results (left panel of Fig.~\ref{fig:final}, Table~\ref{tab:inf-peri}), how long the galaxies in our sample could have remained star-forming. We quantify this by dividing the stellar mass yet to be formed at the first pericentric passage (based on the median estimates in Table~\ref{tab:inf-peri}, last column) by either the minimum ``active'' SFR corresponding to a specific star formation rate (SSFR) of $10^{-11}\,\mathrm{yr}^{-1}$, or a representative SSFR for  an actively star forming galaxy at $z\sim 0.25$ (the typical pericenter time for our sample) of $3\times 10^{-10}\,\mathrm{yr}^{-1}$ \citep{Madau_2014}. The resulting values (Table~\ref{tab:tq}) suggest that star formation typically ceases well under a gigayear after first pericenter in most cases, even in the contrived scenario where it occurs steadily at the threshold SFR for active star formation. We note, however, that the uncertainty in the fraction of stellar mass formed by pericenter for our galaxies is substantial, which prevents our ruling out much longer quenching timescales with confidence.
    
    This serves to highlight that our measurements lead to a qualitatively different picture from most other low-redshift determinations of the quenching timescale. For instance, \citet{Wetzel_2013,Rhee_2020,Oman2021}  suggest that star formation in satellites of comparable mass to those in our sample continues for several gigayears after the first pericentric approach, and that quenching is usually not complete until about the first apocentric passage.
    
    We have not explicitly measured the quenching timescale in this study, so we have to be careful while comparing the quenching scenario obtained in our results with others. For example, \citet{Oman2021} is sensitive to quenching timescale for galaxies undergoing quenching now (i.e., those that had their first pericentric passage $\sim t_\mathrm{q}$ ago -- they measure $t_\mathrm{q}$ with reference to this pericentric time). \citet{Rhee_2020} give a $z=0$ value assuming $t_\mathrm{q}\sim(1+z_\mathrm{inf})^{-1.5}$. The method of \citet{Wetzel_2013} models the full SSFR distribution, and so likely mixes galaxies being quenched at different times with a nontrivial weighting. However, we have estimated infall redshifts of $z \sim 0.6$ for the galaxies in our sample, so we expect to be sensitive to $t_\mathrm{q}$ for satellites that fell in around this time. We therefore expect to find shorter quenching times ($t_\mathrm{q}$ is shorter at higher redshift in comparison to lower redshift -- see Sect.~\ref{sec:intro}) than determinations at $z\sim 0$, but it is not clear by how much. For instance, the multiplicative scaling of $t_\mathrm{q}(z)$ assumed by \citet{Rhee_2020} is ambiguous, as it is sensitive to the definition of the time zero-point. Their reference time is that of first infall across $r=2R_\mathrm{vir}$. Adopting instead the first pericenter reference time of \citet{Oman2021}, for instance, would lead to a qualitatively similar, but quantitatively different prediction. 

    \begin{figure}
        \includegraphics[width=\columnwidth]{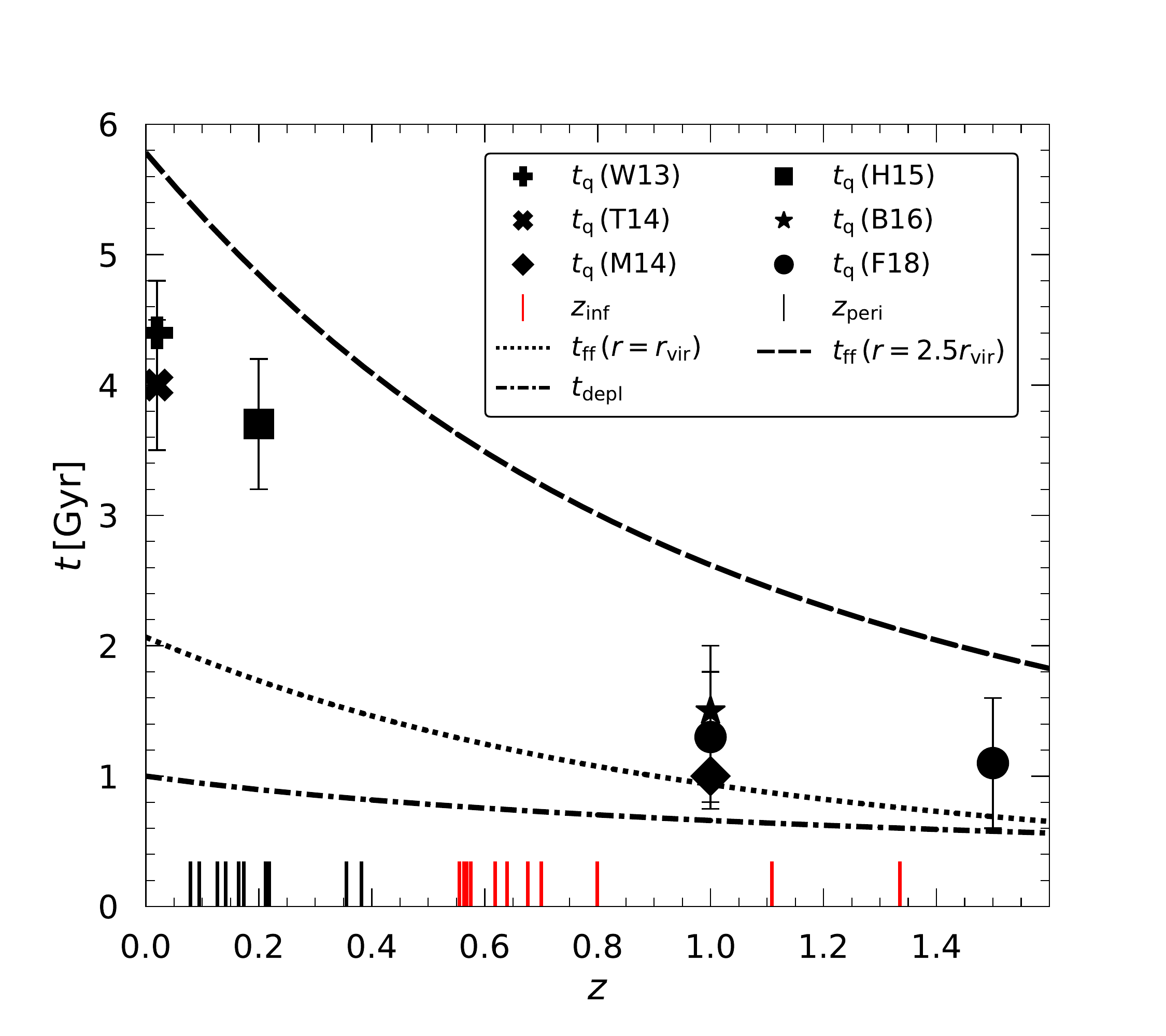}
    \caption[]{Redshift dependence of the quenching timescale $t_\mathrm{q}$. The solid black points show the quenching timescales in clusters at different redshifts as measured by \citet[W13]{Wetzel_2013}, \citet[T14]{taranu2014}, \citet[M14]{Muzzin_2014}, \citet[H15]{Haines_2015}, \citet[B16]{balogh2016evidence}, \citet[F18]{foltz2018evolution}. The dashed and dotted lines show the free fall timescales $t_\mathrm{ff}$ from $r=2.5r_\mathrm{vir}$ and $r=r_\mathrm{vir}$, respectively, for the Coma cluster. The dash-dotted line shows the molecular gas depletion timescale ($t_\mathrm{depl}$) with redshift \citep{tacconi2018phibss}, normalized to $\sim 1$\,Gyr at $z=0$. The red and black ticks show the expected infall and pericenter redshifts, respectively, for the galaxies in this study.}
    \label{fig:tq}
    \end{figure}
    
    We therefore turn to higher-redshift determinations of the quenching timescale for guidance. We compile the quenching timescales ($t_\mathrm{q}$) at different redshifts from several studies and show them in Fig.~\ref{fig:tq} as solid points. We note that the quenching timescales measured at higher redshift \citep[$1.0 \lesssim z \lesssim 1.5$;][]{Muzzin_2014,balogh2016evidence,foltz2018evolution} are much shorter, by $\sim 3$\,Gyr, than lower-redshift ($0 \lesssim z \lesssim 0.2$) determinations \citep{Wetzel_2013,taranu2014,Haines_2015}. Given that the galaxies in our sample fell in around $z\sim 0.6$ (Fig.~\ref{fig:tq}), we expect our results to be consistent with a timescale bounded between the higher and lower redshift measurements.
    
    We also show in Fig.~\ref{fig:tq} the redshift evolution of the free-fall timescale $t_\mathrm{ff}=0.5/\sqrt{G\overline{\rho}(r, z)}$ of the Coma cluster, where $\overline{\rho}(r,z)$ is the mean enclosed density within radius $r$ at redshift $z$, from two different starting radii, $r=r_\mathrm{vir}$ (dotted line) and $r=2.5r_\mathrm{vir}$ (dashed line\footnote{For simplicity, we approximate the free-fall time from $2.5r_\mathrm{vir}$ by computing the mass enclosed within $2.5r_\mathrm{vir}$ at $z=0$ assuming an NFW density profile with $c_\mathrm{vir}=r_\mathrm{vir}/r_s=5.5$, where $r_s$ is the halo ``scale radius,'' and then assuming that the ratio $\overline{\rho}(2.5r_\mathrm{vir}, z)/\overline{\rho}(r_\mathrm{vir},z)$ is a constant over the redshift range of interest. This choice obviates the need to assume a mass accretion history to carry out the calculaion.}). Finally, we show with a dash-dotted line the evolution of the molecular gas depletion timescale with redshift, 
    $$t_\mathrm{depl} \sim (1+z)^{-0.6} (\delta_\mathrm{MS})^{-0.44},$$
    taken from \citet{tacconi2018phibss}. They estimate the molecular gas mass by measuring CO line fluxes of a large sample of star-forming galaxies in the PHIBSS survey between $0 < z < 4$. The depletion timescale depends on the offset in SFR from the star forming main sequence $\delta_\mathrm{MS}=\mathrm{SFR}/\mathrm{SFR}_\mathrm{MS}(M_\star)$; we plot the curve corresponding to $\delta_\mathrm{MS}=0$. This gives an indicative trend, and in any case the uncertain absolute calibration of our STECKMAP SFHs means that we can do little more than assume that our sample of galaxies had typical SFRs around the time they fell into Coma.
    
    We note that $t_\mathrm{depl}$ and $t_\mathrm{ff}$ ($r=r_\mathrm{vir}$) are similar at higher redshift, including at $z \sim 0.6$, which corresponds to the expected infall times for our sample of satellites. The accretion of fresh gas for star formation in cluster satellites is cut off somewhere between $r \sim 2.5r_\mathrm{vir}$ and $r_\mathrm{vir}$, likely toward the latter, and from here the gas reservoir depletes without replenishment. For gas depletion to lead to quenching ($\mathrm{SSFR}<10^{-11}\,\mathrm{yr}^{-1}$) likely takes a few times $t_\mathrm{depl}$ as this timescale is that of an approximately exponential decline in the gas content\footnote{That is, the plausible assumption that the SFR is proportional to the remaining gas supply leads immediately to a relation of the form $M_\mathrm{gas}(t)\propto e^{-t/t_\mathrm{depl}}$.}. At $z\sim 1$, the quenching times $t_\mathrm{q} \sim t_\mathrm{depl}$ therefore seem somewhat too short to be consistent with a starvation scenario, suggesting that some gas removal by other mechanisms is likely required. This hypothesis is further reinforced by the fact that $t_\mathrm{q}\sim t_\mathrm{ff}$, such that these galaxies are quenching around the time when gas removal by ram pressure or tidal stripping is most likely.
    
    In contrast, at low redshift ($z \sim 0$--$0.2$),  the quenching timescale is significantly longer than both the free-fall time (from $\sim r_\mathrm{vir}$), and the depletion timescale. This suggests that recently infalling satellites survive their first pericentric passage with a substantial gas reservoir, enough to sustain SF for several gigayears more, and that they quench after a few gas depletion timescales. This leads to a qualitatively different picture than at $z\sim 1$: $z\sim 0$ satellites seem to undergo starvation-driven quenching, with most of their gas supply being consumed by star formation rather than being stripped away.
    
    As our sample of galaxies correspond to a time intermediate between the observations constraining the two scenarios discussed above, it is interesting to consider which one(s) they may be consistent with. Our finding that sustained star formation after first pericenter is unlikely for our sample of satellites is reminiscent of the $z\sim 1$ ``stripping-assisted'' quenching outlined above. We therefore suggest that the transition to $z\sim 0$ starvation-driven quenching occurs at $z < 0.6$. A key caveat to this argument is that if satellites in our sample had their supply of fresh gas cut off much earlier than around when they entered the Coma virial radius (e.g., if they were preprocessed in infalling groups), they may then plausibly quench by starvation by the time of their first pericentric passage in Coma. \citet{bahe2015}  reported a similar prediction from simulations that predict a transition from stripping to consumption-driven quenching progressing from higher to lower redshift. Disentangling these possibilities will benefit from larger samples of satellites with resolved SFHs to help strengthen (or rule out) the tentative link we find between the quenching time and the first pericentric passage.

    In the right panel of Fig.~\ref{fig:final}, we highlight that higher mass galaxies ($\log (M_\star/\mathrm{M}_\odot) > 10$) have likely formed a lower fraction of their stellar mass ($\sim 5\%$) between infall and pericenter than lower mass galaxies, which form almost double ($\sim 10\%$) the fractional amount. Furthermore, all of the galaxies in our sample have nearly completed their star formation around the pericenter (see left panel of Fig.~\ref{fig:final}). This could simply be a reflection of the somewhat higher SSFRs of lower mass galaxies \citep[see Fig.~5 of][the SSFR evolution with redshift up to $z \sim 1.4$ for galaxies in mass range of $9.5 \leqslant \log(M_\star/\mathrm{M}_\odot) \leqslant 11.5$]{Ilbert2015}. Alternately, this could be interpreted in terms of the quenching timescale, suggesting that massive satellites quench earlier than their lower mass counterparts. We note that this interpretation is at least qualitatively consistent\footnote{It is also consistent with that of \citet{Oman_2016}, but this analysis makes erroneous assumptions affecting the measurements, as pointed out by \citet{Oman2021}; the latter work should be taken as superseding the former.} with those of \citet{Wetzel_2013,Rhee_2020}, who also suggest a shorter quenching timescale for massive satellites \citep[see also][who found similar results for cluster satellites at higher redshift]{Contini_2020}. 
    
    The infall time probability distribution for GMP~3254, shown in the left panel of Fig.~\ref{fig:cdf-pdf}, is bimodal. This begs the question of whether our simple treatment of the distribution, characterizing it by its width, biases our conclusions. We have examined the case of GMP~3254 in more detail by decomposing the distribution into an earlier (older) and a later (younger) peak, and we have determined the percentage of stellar mass formed at the expected infall time of each peak. We find that the earlier peak is consistent with our overall qualitative conclusions drawn from the full distribution, while the later peak, considered alone, may hint that the galaxy was quenched -- or nearly quenched -- before it entered the cluster. However, extending this analysis to our entire sample is prone to a two caveats: First, we do not observe a consistent bimodal trend in PDFs of all the galaxies; second, the conclusions are not highly sensitive to our choice of ignoring the bimodality, and we have no clear reason to prefer one peak over the other. In addition, the cumulative distribution shown in the right panel of Fig.~\ref{fig:cdf-pdf} explicitly accounts for the correlation across the two probability distributions.
 
    Finally, we note that many Coma cluster galaxies in our study might have spent time in overdense regions prior to infall: At least some of them probably spent some time as satellites of infalling groups. This may be an interesting topic for future work as the approach we have adopted applied to larger samples may be able to find evidence for (or against) two distinct populations of satellites, corresponding to preprocessed and direct infall objects.

\subsection{The stellar-to-halo mass relation}\label{sec:disc-shmr}
    
    The \citet{Behroozi_2010} SHMR (see Sect.~\ref{sec:obsorbits}) uses subhalo abundance matching (SHAM) to assign stellar masses to subhalos. The SHAM technique is very successful in matching observed galaxy statistics in spite of the differing evolutionary histories of satellite and central galaxies. There are several ways in which SHAM can be implemented; one particular way is based on the subhalo mass at the current epoch. However, since satellite subhalos are in general stripped of a fraction of their halo mass, before the stellar mass begins to be stripped, this introduces a bias for satellites. \citet[][see their Appendix~A]{Wetzel_2013} argue that $M_{\mathrm{max}}$ is better correlated with the stellar mass of subhalos since it always occurs before infall: It is not altered during satellite orbits. In our study, we have implemented SHAM based on $M_{\mathrm{max}}$ rather than the $z=0$ $M_{\mathrm{sat}}$.

    The use of this relation relies on the assumptions that (i) the satellites have the same stellar mass now as they did at infall, in other words tidal stripping of stars and stellar mass growth through star formation are negligible, and (ii) the SHMR has not evolved significantly since the time when our sample of galaxies became satellites. Based on Table~\ref{tab:inf-peri}, the satellites have grown by up to about $25\%$ (or $70\%$ at $68\%$ confidence) since infall. This corresponds to $\sim 0.1\,\mathrm{dex}$ ($0.2\,\mathrm{dex}$ at $68\%$ confidence). While this is several times smaller than the $0.5\,\mathrm{dex}$ interval in halo mass that we use to match simulated and observed satellites, the bias introduced is clearly systematic, causing all the halo masses to be overestimated.
    
    The degree to which our sample of satellite galaxies may have been tidally stripped of stars is more difficult to assess; however, we think it unlikely that any satellites in our sample are ``heavily'' stripped of stars. This is first because the stellar portion of the galaxies are the most tightly bound, so that once the system begins to lose substantial amounts of stellar mass, it is already close to being completely disrupted \citep{Bahe_2019}. This makes satellites with significant tidal stellar mass loss short-lived, so finding multiple examples in our sample is unlikely. Second, our satellites have a distribution of shapes typical of central ETGs with similar masses, with ellipticity ($e=1-b/a$) in the range $0.2$--$0.5$ \citep{hoyos2011}, while significantly stripped satellites are expected to be noticeably rounder, on average \citep[e.g.,][]{Barber_2014}. We also note that any hypothetical stellar mass loss to tides would partially compensate the stellar mass growth reflected in the SFHs.
    
    The SHMR is redshift dependent. For our sample of satellites, using the $z=1$ SHMR instead of that at low redshift ($z=0.1$) -- the bulk of our sample have likely infall times around $z\sim 0.6$ -- would cause us to revise their halo masses upward by up to $0.2\,\mathrm{dex}$ \citep{Behroozi_2010}. We note that this does not linearly combine with the systematic offset due to their stellar mass growth, described above since when using a higher redshift SHMR, the correct stellar mass to use is that ``at the same redshift''. However, the biases introduced in our analysis by these two effects act in opposite senses, mitigating their overall severity.
    
    We chose to use the \citet{Behroozi_2010} SHMR, but there are many other published relations that could have been used in its stead. The \citet{Behroozi_2010} relation is convenient in that it lies approximately in the middle of the scatter between the various proposed relations \citep[e.g., the compilation in][ their Fig.~34]{Behroozi2019}. However, the SHMR does include a $\sim 0.15\,\mathrm{dex}$ scatter, which we have ignored; we find it unlikely that this additional scatter will make a significant change to our results. At the stellar mass scale of interest, the (minimum-to-maximum) scatter between the various proposed mean relations is about $0.3\,\mathrm{dex}$. This represents a possible systematic bias affecting our analysis; unlike the others discussed in this subsection, the sign of this bias is unknown.
    
    Based on the above discussion, we (conservatively) estimate that there may be up to a $\sim 0.5\,\mathrm{dex}$ systematic mismatch between the actual halo masses at infall of our sample of Coma satellites and those of the satellite halos selected from the simulations.
    
\subsection{Systematic errors}\label{sec:systematic}
    \begin{figure*}
        \includegraphics[width=\textwidth]{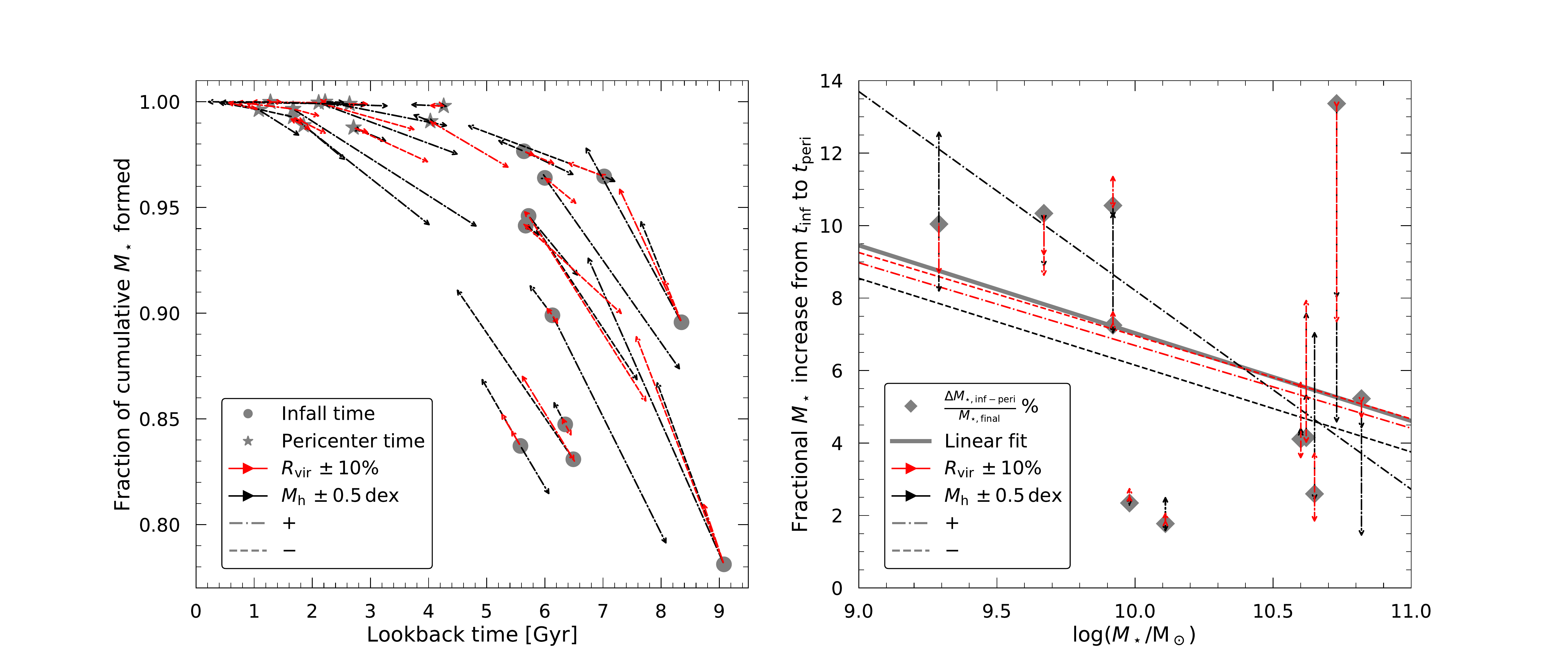}
    \caption[]{Estimation of systematic errors. 
    \emph{Left panel}: Change in fractional stellar mass formed at infall and pericenter times incurred by introducing a systematic $0.5\,\mathrm{dex}$ offset in the satellite halo mass (black arrows) or a 10\% change to the Coma virial radius, propagated to all quantities that depend on this measurement (red arrows). Dash-dotted (dashed) arrows indicate the result of a positive (negative) shift in the relevant quantity. These shifts are typically small and hence do not affect the overall qualitative interpretation of our work. We note that the axes of the figure are zoomed-in relative to Fig.~\ref{fig:final} to show the very small shifts. See Sect.~\ref{sec:systematic} for further details.
    \emph{Right panel}: Change in fractional stellar mass formed between infall and pericenter due to the introduction of the same systematic offsets as in the left panel.}
    \label{fig:zoom}
    \end{figure*}

    We checked whether such a systematic bias could influence our qualitative interpretation of Fig.~\ref{fig:final} by increasing all the halo mass estimates for our sample of satellites (Table~\ref{tab:orb_prop}) by $0.5\,\mathrm{dex}$ and repeating the rest of the analysis. The resulting shifts of the points in Fig.~\ref{fig:final} are illustrated in Fig.~\ref{fig:zoom} with dash-dotted black arrows. We repeated the same exercise but instead reduced all halo mass estimates by $0.5\,\mathrm{dex}$; the resulting shifts are shown with black dashed arrows in Fig.~\ref{fig:zoom}. While many individual measurements move significantly, the overall qualitative picture is unchanged: These satellites form up to $\sim 15\%$ of their stellar mass between infall and their first pericentric passage, and very little thereafter.

    We also checked the sensitivity of our results for another possible bias, unrelated to those discussed in Sect.~\ref{sec:disc-shmr}, by following a similar procedure. We adjusted the virial radius of the Coma cluster ($R_{\mathrm{vir,Coma}}$) by $\pm10\%$, consistent with the plausible interval of $2.6$--$2.9\,\mathrm{Mpc}$ given in \citet{Lokas_2003}, and propagated the change to $M_{\mathrm{vir,Coma}}$ and $\sigma_{3{\mathrm{D,Coma}}}$. We then repeated the selection of host systems from the N-body simulation, made new selections of satellite halos, and repeated the determination of the infall and pericenter times and the corresponding stellar mass fractions formed at those times. The relative shifts of the points in Fig.~\ref{fig:final} are again shown in Fig.~\ref{fig:zoom}, with that due to an increased (decreased) $R_{\mathrm{vir,Coma}}$ shown with a dash-dotted (dashed) red arrow. We find again that, although some points may move substantially, the overall interpretation is robust against moderate uncertainty in the properties of the Coma cluster.
    
    \begin{table}
    \centering
        \caption[]{Properties of SSPs in STECKMAP.}
         \label{tab:ssp}
        \begin{tabular}{l c c c c}
        \hline
        \noalign{\smallskip}
        SSP & $\lambda$\,(\AA) & $\Delta\lambda$\,(\AA) & Age\,(Gyr) & [Z/H] \\
        \noalign{\smallskip}
        \hline
        \noalign{\smallskip} 
        \texttt{BC03} & $3200$--$9500$ & $3.0$ & $0.0001$--$17$ & [$0.3$,$-2.0$]\\
        \texttt{PHR} & $4000$--$6800$ & $2.0$ & $\phantom{00}0.02$--$17$ & [$0.2$,$-2.0$]\\
        \texttt{GD05} & $3000$--$7000$ & $0.3$ & $\phantom{00}0.02$--$17$ & [$0.2$,$-2.0$]\\
        \texttt{MILES} & $3525$--$7500$ & $2.3$ & $\phantom{00}0.02$--$17$ & [$0.2$,$-1.3$]\\
        \noalign{\smallskip}
        \hline
        \end{tabular}
    \tablefoot{The general properties of SSPs available in STECKMAP, namely, \texttt{BC03} \citep{bruzal2003}, \texttt{PHR} \citep{leBorgne2004}, \texttt{GD05} \citep{gonzalez2005}, \texttt{MILES} \citep{Vazdekis_2010}. The table shows wavelength ($\lambda$) range in\,\AA, spectral resolution ($\Delta\lambda$) in\,\AA, age range in\,Gyr, and metallicity ([Z/H]) range.}
    \end{table}
    
    \begin{figure*}
        \includegraphics[width=\textwidth]{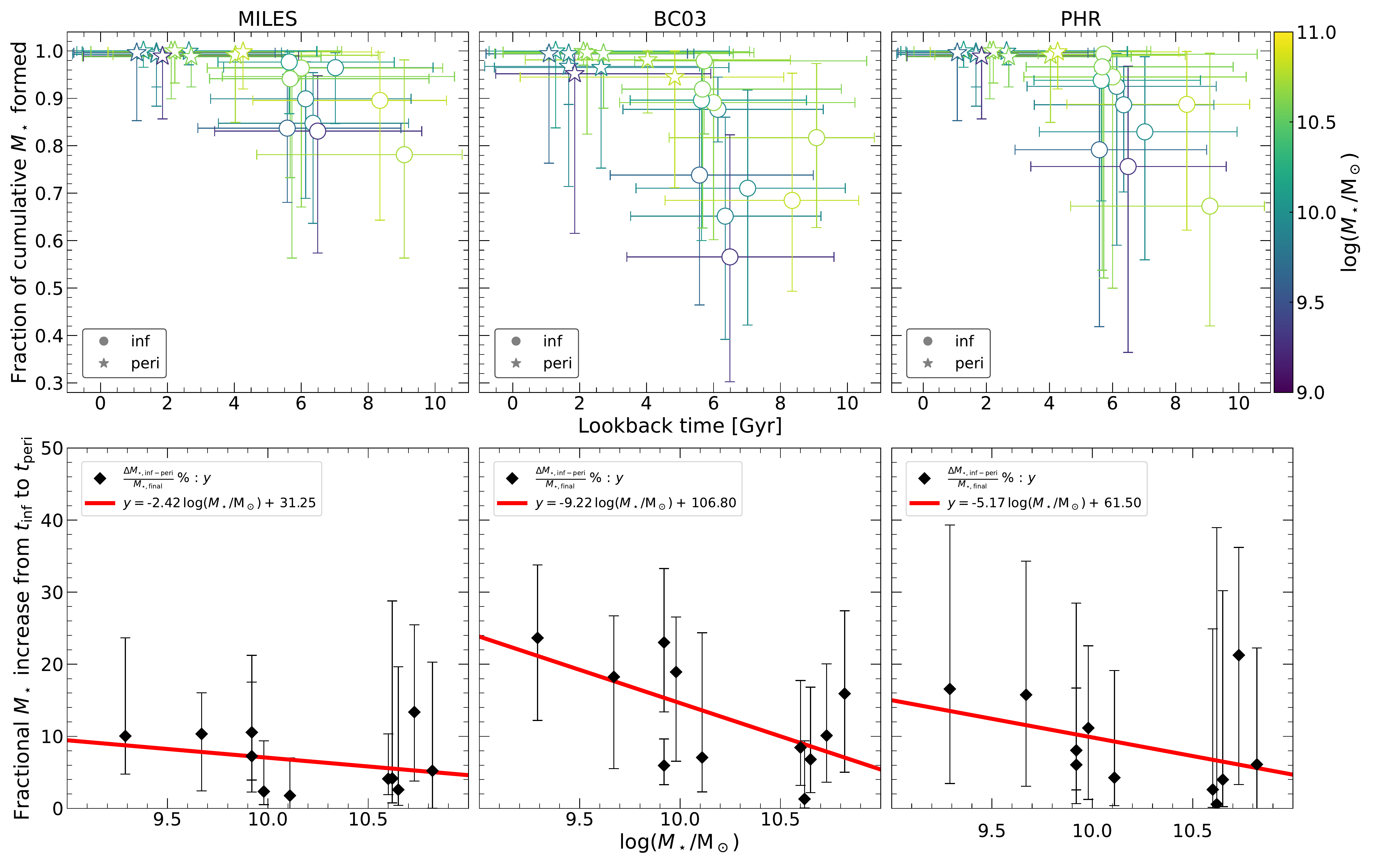}
    \caption[]{Fraction of stellar mass formed at infall and pericenter time (top panel) and increase in fractional stellar mass between infall and pericenter (bottom panel) using different SSP models: \texttt{MILES} (left panel), \texttt{BC03} (center panel), and \texttt{PHR} (right panel) are shown. \texttt{MILES} provides the most robust results, with the smallest error bars, in comparison to the other SSPs.} 
    \label{fig:comp_ssp}
    \end{figure*}
    
    Finally, the fiducial result shown in Fig.~\ref{fig:final} is based on STECKMAP SFHs using the \texttt{MILES} SSP models, as discussed in Sect.~\ref{sec:stellarpops}. STECKMAP provides four options for stellar population libraries: \texttt{BC03} \citep{bruzal2003}, \texttt{MILES} \citep{Vazdekis_2010}, \texttt{PHR} \citep{leBorgne2004} and \texttt{GD05} \citep{gonzalez2005}. Their general properties like spectral resolution and range of wavelength, age, and metallicity are listed in Table~\ref{tab:ssp}. We check the sensitivity of our results to the choice of SSP model by repeating our analysis using the two other SSP models available in STECKMAP: \texttt{BC03} and \texttt{PHR}. Fig.~\ref{fig:comp_ssp} shows the cumulative stellar mass formed at infall and pericenter, and the fraction of stellar mass formed between infall and pericenter, for all three SSP models. The $68\%$ confidence intervals are somewhat wider when \texttt{BC03} or \texttt{PHR} are used, but the results are otherwise very similar. There is a hint in Fig.~\ref{fig:comp_ssp} that using \texttt{MILES} leads to the weakest gradient in the stellar mass formed between infall and pericenter as a function of stellar mass, which mildly increases our confidence that this may be a real trend.

\section{Conclusions}
    
    In this study, we have attempted to understand the environmental quenching of satellites in the high-density environment of galaxy clusters by linking their SFHs with their orbital histories. Our observed sample consists of high-S/N spectra of 11 ETGs \citep{Trager_2008} around the Coma cluster center, from which SFHs were inferred based on the relative SFR values obtained using STECKMAP (see Sect.~\ref{sec:stellarpops}). The intra-cluster orbits of these galaxies have been constrained based on their projected phase space coordinates and a library of orbits extracted from an N-body simulation (see Sects.~\ref{sec:orbits} and \ref{sec:obsorbits}). We match the cumulative distributions of orbital parameters against the satellite SFHs to infer the fraction of stellar mass formed by the expected infall and first pericenter times. Our main results (see Sect.~\ref{subsec:pdf_sfr}, Table~\ref{tab:inf-peri} and Fig.~\ref{fig:final}) are summarized as follows.
    
    \begin{enumerate}
        \item We find that infall across $2.5r_\mathrm{vir}$ for galaxies in our sample occurred around $z \sim 0.6$, and they reached their first pericenter $\sim 4$\,Gyr later.
        \item Massive galaxies ($\log(M_\star/\mathrm{M}_\odot) \gtrsim 10$) in our sample have formed a higher fraction of stellar mass ($\gtrsim 90\%$) before infall than less massive galaxies ($\sim 80$--$90\%$), although the uncertainties are large.
        \item Massive galaxies in the Coma cluster quenched on a shorter timescale than their lower mass counterparts.
        \item Conversely, our sample galaxies formed a significant fraction of stellar mass  ($\gtrsim 5\%$) within the high-density Coma cluster environment. 
        \item  The galaxies in our sample are likely to have formed nearly all of their stellar mass ($\gtrsim 98\%$) by the time of their first pericentric passage. It is therefore likely that these galaxies quenched around (or within $\lesssim 1$\,Gyr after) first pericenter.
    \end{enumerate}
    
    
    In contrast with studies sensitive to the current quenching timescale at $z \sim 0$ that find that star formation in satellites continues well after their first pericentric passage \citep[e.g.,][]{Wetzel_2013,Rhee_2020,Oman2021}, our measurements are consistent with truncation of star formation in present-day quiescent cluster galaxies around first pericenter. This is reminiscent of higher-redshift ($z\sim 1$) determinations of the quenching timescale for satellites \citep[e.g.,][]{Muzzin_2014,balogh2016evidence,foltz2018evolution}. Putting this together with our results, the typical infall redshift for our sample ($z\sim 0.6$), the relevant free-fall and gas consumption timescales and their scalings with redshift, our interpretation is that galaxies in our sample likely lost substantial amounts of gas to ram pressure and/or tidal stripping. Put another way, their quenching was ``not'' primarily starvation-driven. We caution that this interpretation would change if most of the satellites in our sample were satellites of smaller groups before falling into the Coma cluster (preprocessed).
    
    This exploratory analysis demonstrates the information-rich nature of SFHs when combined with orbital information from simulations. Repeating our study with the SFHs replaced with only a present-day star formation indicator (e.g., broadband color) would lead to only trivial conclusions. Our approach paves the way for future work leveraging large, deep spectroscopic surveys of clusters. WEAVE \citep{dalton2014project}, with its $2^\circ$ field of view and nearly 1000 individual fibers, currently being installed on the William Herschel Telescope (WHT), will be an ideal instrument for such a study, and the Infall Regions subsurvey of the WEAVE Clusters survey will determine SFHs of tens of thousands of individual galaxies out to several virial radii in more than a dozen clusters.

\begin{acknowledgements}
      We thank S. Bose and A. Jenkins for providing us the code and initial conditions for the N-body simulation and R. Peletier for a careful reading of an early version of the text. We also thank the anonymous referee for providing valuable (and rapid) feedback on the manuscript. 
      
      KAO acknowledges support by the Netherlands Foundation for Scientific Research (NWO) through VICI grant 016.130.338 to M.~A.~W.~Verheijen, and support by the European Research Council (ERC) through Advanced Investigator grant to C.~S.~Frenk, DMIDAS (GA~786910).
      
      This work used the DiRAC@Durham facility managed by the Institute for Computational Cosmology on behalf of the STFC DiRAC HPC Facility (www.dirac.ac.uk). The equipment was funded by BEIS capital funding via STFC capital grants ST/K00042X/1, ST/P002293/1, ST/R002371/1 and ST/S002502/1, Durham University and STFC operations grant ST/R000832/1. DiRAC is part of the National e-Infrastructure.
\end{acknowledgements}
\bibliographystyle{aa}
\bibliography{ref.bib}
\end{document}